\documentclass{article}

\usepackage{microtype}
\usepackage{graphicx}
\usepackage{subcaption}
\usepackage{booktabs} 

\usepackage{hyperref}

\usepackage[preprint]{icml2026}

\usepackage{amsmath}
\usepackage{amssymb}
\usepackage{mathtools}
\usepackage{amsthm}

\usepackage[capitalize,noabbrev]{cleveref}

\usepackage{makecell}
\usepackage{algorithm}
\usepackage{microtype}
\usepackage{booktabs}
\usepackage{hyperref}
\usepackage{amsmath}
\usepackage{amssymb}
\usepackage{mathtools}
\usepackage{amsthm}
\usepackage{url}
\usepackage{colortbl}
\usepackage{listings}
\usepackage[dvipsnames]{xcolor}
\usepackage{wrapfig}
\usepackage{graphicx}
\usepackage{subcaption}
\usepackage{xspace}
\usepackage{multicol, multirow, bigstrut}
\usepackage{verbatim}
\usepackage{mdframed}
\usepackage{adjustbox}
\usepackage{tabularx}
\usepackage{enumitem}
\usepackage{xurl}  

\usepackage[most]{tcolorbox}
\definecolor{promptbgcolor}{RGB}{245,245,245} 
\definecolor{promptframecolor}{RGB}{100,100,100} 
\definecolor{prompttitlebg}{RGB}{220,220,220} 
\newtcolorbox[auto counter, number within=section]{promptbox}[2][]{
    enhanced,
    breakable, 
    fonttitle=\bfseries,
    title={Prompt \thetcbcounter: #2}, 
    colback=promptbgcolor, 
    colframe=promptframecolor, 
    coltitle=black, 
    colbacktitle=prompttitlebg, 
    attach boxed title to top left={xshift=5mm, yshift=-3mm}, 
    boxed title style={colframe=promptframecolor, colback=prompttitlebg},
    sharp corners=downhill, 
    arc=3mm,
    boxrule=0.8pt,
    left=5mm, right=5mm, top=5mm, bottom=3mm, 
    #1
}

\newcommand{\role}[1]{\textbf{\textcolor{blue!60!black}{\textsc{#1}:}}\quad}
\newcommand{\placeholder}[1]{\textcolor{red!60!black}{\texttt{[#1]}}}

\tcbuselibrary{listings}
\definecolor{Gray}{gray}{0.9}
\definecolor{ForestGreen}{rgb}{0.13, 0.55, 0.13} %
\newcommand{\our}[1]{\textsc{AdverMCTS}}

\theoremstyle{plain}

\theoremstyle{definition}

\theoremstyle{remark}

\usepackage[disable,textsize=tiny]{todonotes}

\icmltitlerunning{\our{}}

\begin{document}

\twocolumn[
  \icmltitle{AdverMCTS: Combating Pseudo-Correctness in Code Generation via Adversarial Monte Carlo Tree Search}



  \icmlsetsymbol{equal}{*}

  \begin{icmlauthorlist}
    \icmlauthor{Qingyao Li}{sjtu}
    \icmlauthor{Weiwen Liu}{sjtu}
    \icmlauthor{Weinan Zhang}{sjtu}
    \icmlauthor{Yong Yu}{sjtu}
    \icmlauthor{Bo An}{ntu}
  \end{icmlauthorlist}

  \icmlaffiliation{sjtu}{Shanghai Jiao Tong University, Shanghai, China}
  \icmlaffiliation{ntu}{Nanyang Technological University, Singapore}
  \icmlcorrespondingauthor{Qingyao Li}{ly890306@sjtu.edu.cn}
  \icmlcorrespondingauthor{Weinan Zhang}{wnzhang@sjtu.edu.cn}
  \icmlcorrespondingauthor{Bo An}{boan@ntu.edu.sg}

  \icmlkeywords{Large Language Models, Code Generation, Monte Carlo Tree Search, Adversarial Search, Test-Time Scaling}

  \vskip 0.3in
]



\printAffiliationsAndNotice{}  

\begin{abstract}
Recent advancements in Large Language Models (LLMs) have successfully employed search-based strategies to enhance code generation. However, existing methods typically rely on static, sparse public test cases for verification, leading to \textit{pseudo-correctness}—where solutions overfit the visible public tests but fail to generalize to hidden test cases. We argue that optimizing against a fixed, weak environment inherently limits robustness. To address this, we propose \our{}, a novel adversarial Monte Carlo Tree Search framework that combats \textit{pseudo-correctness} by coupling code search with active vulnerability discovery. \our{} formulates generation as a minimax-style game between a \textit{Solver} agent, which synthesizes code candidates, and an \textit{Attacker} agent, which evolves to generate targeted corner test cases that exploit logical divergences in the current code pool. These discovered tests form a dynamic, progressively hostile filter that penalizes fragile reasoning. Extensive experiments demonstrate that \our{} significantly outperforms state-of-the-art baselines, effectively reducing false positive rates and forcing the model to generalize beyond the initial constraints. The resources of this work are available at \href{https://anonymous.4open.science/r/AdverMCTS_open-A255}{https://anonymous.4open.science/r/AdverMCTS-A255}.
\end{abstract}

\section{Introduction}
The advent of Large Language Models (LLMs) has fundamentally transformed the landscape of automated code generation~\citep{odeh2024comparative, jiang2024survey, yang2025empirical, dong2025survey}, enabling systems to solve standard programming tasks with remarkable proficiency \citep{chen2021evaluating, li2022competition,wang2025teaching}. However, as the focus shifts towards complex, competition-level problems that demand deep algorithmic reasoning, the paradigm is evolving from simple ``System 1'' token prediction to ``System 2'' deliberation~\citep{xiang2025towards, li2025system}, often characterized as Test-Time Compute (TTC) scaling \citep{brown2024large, snell2024scaling}. Central to this shift is the reliance on execution-based verification~\citep{zhong2024ldb, ni2023lever, dong2025codescore}, where the correctness of a generated code is inferred from its behavior on a set of test cases. In this context, the interplay between code synthesis and test validation has become a pivotal axis for advancing model performance.

\begin{figure}[t]
    \centering
    \includegraphics[width=0.97\linewidth]{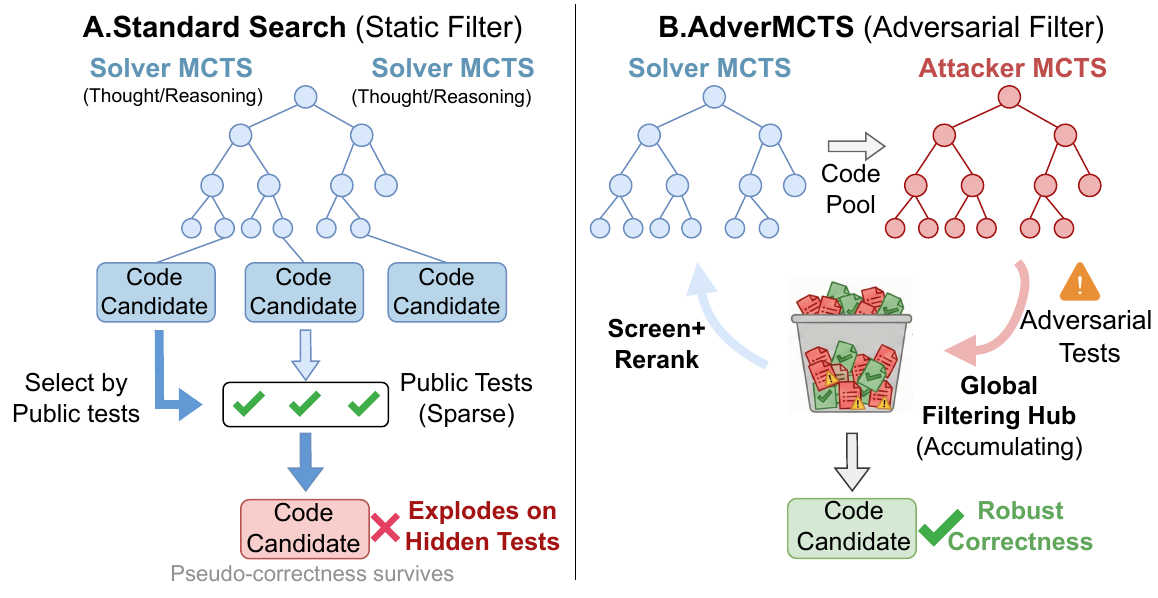}
    \caption{Conceptual Comparison. (A) Standard Search relies on sparse public tests, creating a ``leaky'' filter prone to pseudo-correctness. (B) \our{} employs an active Attacker to co-evolve a progressively stricter environment, exposing hidden bugs and enforcing robust correctness.}
    \vspace{-18pt}
    \label{fig:intro}
\end{figure}

To leverage this verification signal, recent research has introduced various search-guided frameworks that structure decoding as a navigation problem~\citep{chen2024code, gou2024rrgcode, lyu2025let}. Prominent approaches such as Tree of Thoughts (ToT) \citep{yao2023tree} and Language Agent Tree Search (LATS) \citep{zhou2023language} integrate planning with Breadth First Search~\citep{kurant2010bias} and Monte Carlo Tree Search (MCTS)~\citep{browne2012survey}, using the LLM as both a policy and a value estimator. Similarly, methods like PG-TD \citep{zhang2023planning} and CodeT \citep{chen2022codet} utilize rollout execution on sample tests to guide the generation trajectory. These search-based strategies have successfully pushed the boundaries of code generation, achieving significant performance gains on benchmarks by enabling lookahead and backtracking capabilities that are absent in greedy decoding~\citep{li2025codetree}.

Despite these advancements, we argue that a key challenge is still under-addressed: \textit{pseudo-correctness}—generating solutions that overfit the public tests while failing on the underlying logic required by the hidden test suite. 
Public tests typically sample from simplified sanity checks, whereas hidden tests probe the long-tail of corner cases. Consequently, static verification signals are frequently insufficient to expose hidden bugs, and even sophisticated search can be misled into preferring fragile code, creating a \emph{survivorship bias} where many ``surviving'' candidates are merely overfitted solutions. We empirically verify this bottleneck in Appendix ~\ref{app:emp_pesudo_correct}, confirming that while search spaces often contain correct solutions, sparse public tests fail to identify them. Thus, the bottleneck is not the Solver's capacity to generate correct solutions, but the environment's capacity to \emph{discriminate} them at inference time. This calls for a mechanism that actively strengthens verification, rather than merely expanding the candidate set.

This paper presents a unique perspective on the problem: robust code generation should be viewed as an adversarial game between a solver that proposes solutions and an attacker that actively searches for failure-inducing tests. Intuitively, if the search procedure is only navigating within a fixed, weak environment (the public tests), then it is optimizing the wrong objective. What we need at test time is a mechanism that co-evolves solutions and constraints: as the solver improves, the environment should become more hostile, continuously surfacing new corner cases that invalidate pseudo-correct codes. Figure ~\ref{fig:intro} illustrates this paradigm shift: while standard search is limited by static verification, our approach constructs a dynamic hostile environment to mitigate survivorship bias.
 
Building on this perspective, we propose \our{}, an adversarial Monte Carlo tree search framework that addresses pseudo-correctness by coupling solution search with vulnerability discovery. \our{} instantiates a minimax-style interaction between two agents: (i) a Solver MCTS that searches over chain-of-thought trajectories and continuously synthesizes executable candidate codes during simulation, and (ii) an Attacker MCTS that conditions on the evolving code pool and searches for discriminative tests via divergence-driven test synthesis. A global test filter adjudicates and accumulates valid corner cases into a dynamic test suite, which is then used as hard constraints for screening and final re-ranking codes---forcing the solver to generalize beyond the initial public tests. Our contributions are summarized as follows:
\begin{itemize}[leftmargin=*, topsep=0pt, partopsep=0pt, itemsep=0pt, parsep=2pt]
    \item We propose \our{}, a dual-agent search framework that enhances code robustness through iterative adversarial interaction. To the best of our knowledge, \our{} is the first to unify code search with an active adversarial test search at test time to tackle competition-level programming problems.
    \item We develop key mechanisms that make adversarial test-time search effective in practice, including a persistent attacker tree, divergence-driven test synthesis, and a global test filter that turns discovered corner cases into reusable hard constraints.
    \item Extensive experiments demonstrate that \our{} significantly outperforms state-of-the-art baselines. Notably, our analysis reveals that the adversarial pressure effectively reduces the false positive rate of generated codes, validating the efficacy of ``hostile'' supervision in test-sparse environments.
\end{itemize}

\section{Related Work}
\paragraph{Competition-level Code Generation}
Early LLMs for code such as OpenAI Codex~\citep{chen2021evaluating} achieved strong results on standard benchmarks~\citep{zhou2023language}, yet remained challenged by competition-level problems that require deeper algorithmic reasoning~\citep{li2023starcoder, lozhkov2024starcoder, hui2024qwen2, paul2024benchmarks}. AlphaCode~\citep{li2022competition} marked the first non-trivial breakthrough in competitive programming, reaching median competitor-level performance on Codeforces~\citep{mirzayanov2020codeforces}. Building on this, a growing line of work improves reliability via search-guided generation~\citep{princis2025treecoder, jiang2024self, wang2024planning, chen2024tree, gao2024search, li2025codeprm, li2025rethinkmcts}. Representative examples include PG-TD~\citep{zhang2023planning}, which executes candidate programs on sample tests for lookahead planning; CodeT~\citep{chen2022codet}, which generates additional tests to select solutions and notably improves pass@1 on HumanEval~\citep{li2024humaneval}; and LATS~\citep{zhou2023language}, which integrates MCTS into code generation with the LLM as the policy. While these methods allocate computation to sampling or single-agent search over code, our approach instead introduces an adversarial MCTS framework that couples a code-generating Solver with a test-generating Attacker, enabling targeted failure discovery and more robust selection beyond prior single-agent search paradigms.

\paragraph{Test Time Computing Scaling}
Test-Time Computing (TTC) frames inference-time compute as a key lever for improving model performance~\citep{muennighoff2025s1, chen2024expanding, zhang2025survey, li2025s, zeng2025revisiting}, enabling deeper ``System~2'' style reasoning \citep{brown2024large, snell2024scaling}. This motivates search-structured decoding, where methods such as Tree of Thoughts (ToT) \citep{yao2023tree} and Reasoning via Planning (RAP) \citep{hao2023reasoning} extend Chain-of-Thought \citep{wei2022chain} with lookahead/backtracking, and RethinkMCTS \citep{li2025rethinkmcts} further repairs erroneous nodes using fine-grained feedback.
However, these approaches largely verify against static signals (e.g., LATS or RethinkMCTS), similar in spirit to fixed-signal self-improvement such as Reflexion \citep{shinn2023reflexion}, and differ from adversarial training efforts like ATGen \citep{li2025atgen}. \our{} instead introduces an active adversary at test time: an Attacker that adaptively hardens the verification environment, forcing the Solver to generalize beyond the initial constraints.

\section{\our{}}
\label{sec:method}

\subsection{Problem Formulation}
Given a natural language problem description $P$, robust code generation aims to synthesize a program $C$ that satisfies the underlying semantics of $P$.
Correctness is evaluated by executing $C$ on a test suite $\mathcal{T}$, which is split into a small set of public tests $\mathcal{T}_{\text{pub}}$ and a larger hidden set $\mathcal{T}_{\text{hidden}}$.
At inference time, the solver observes only $(P, \mathcal{T}_{\text{pub}})$, where $\mathcal{T}_{\text{pub}}$ provides limited verification and may fail to expose corner cases.
Consequently, passing $\mathcal{T}_{\text{pub}}$ does not guarantee generalization to $\mathcal{T}_{\text{hidden}}$.
Our goal is to learn a policy that, using only $(P, \mathcal{T}_{\text{pub}})$, produces a solution $C^*$ that maximizes performance on $\mathcal{T}_{\text{hidden}}$.

\subsection{Overview}
\begin{figure*}
    \centering
    \includegraphics[width=0.97\linewidth]{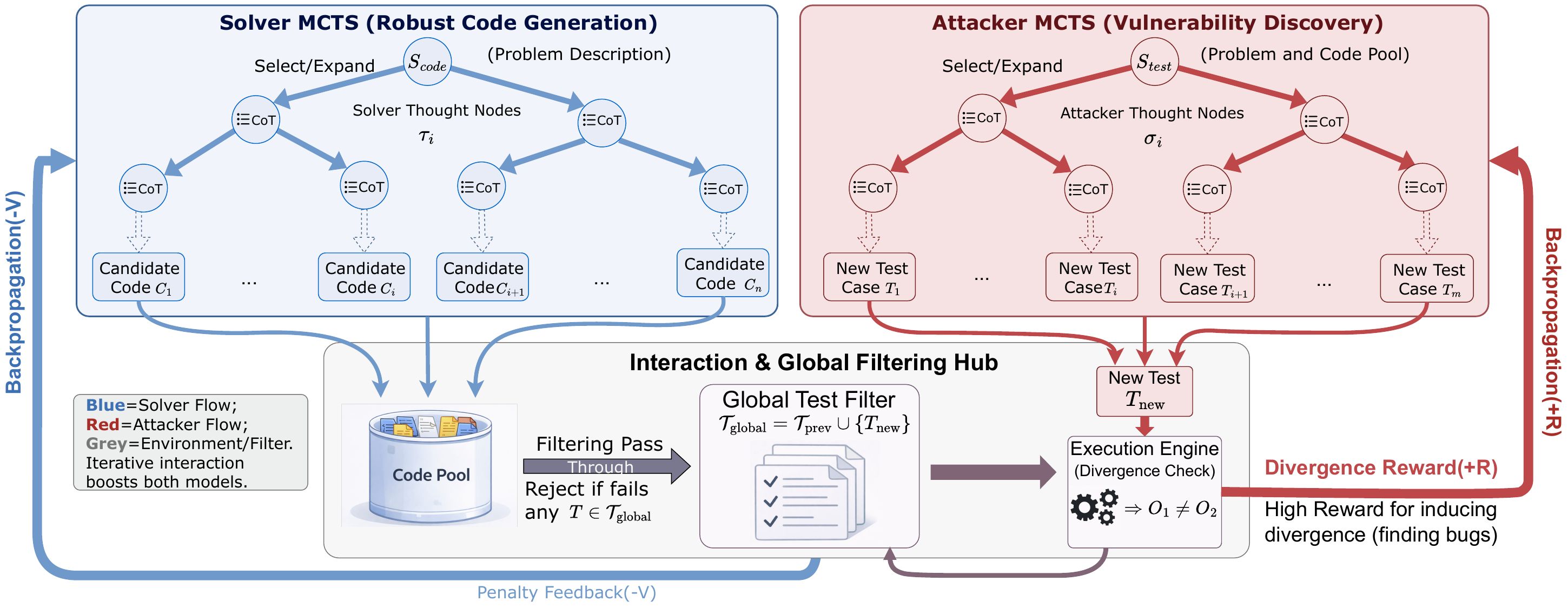}
    \caption{Overview of \our{}. A minimax interaction where the Solver (blue) generates code and the Attacker (red) synthesizes adversarial tests. The Global Hub turns valid attacks into constraints. Feedback is dual: divergence rewards (+R) for the Attacker and penalties (-V) for the Solver enforce robust generalization.}
    \vspace{-5mm}
    \label{fig:overview}
\end{figure*}

We propose \our{}, an adversarial MCTS framework to enhance code generation robustness. As illustrated in Figure \ref{fig:overview}, the framework operates as an iterative game between two agents: a \emph{Solver} and an \emph{Attacker}. The \emph{Solver} aims to synthesize robust code that satisfies the problem requirements, while the \emph{Attacker} actively searches for ``vulnerability'' test cases that induce behavioral divergence in the \emph{Solver}'s generated code. The two agents interact through a shared \emph{Code Pool} and a dynamic \emph{Global Test Filter}, iteratively refining both the solution and test quality. The detailed pseudocode is provided in Algorithm~\ref{alg:advermcts} in the Appendix~\ref{ag:algs}.

\subsection{Solver MCTS: Robust Code Generation}
\label{subsec:solver}

The Solver aims to synthesize a robust program $C$ given the problem description $P$. The search tree acts as a structured reasoning space, where the root represents the initial state $s_0 = {P}$, and each node $s_\text{t} = \{P, \tau_1, \dots, \tau_t\}$ represents a partial solution consisting of a sequence of Chain-of-Thought (CoT) steps. The Solver iteratively builds the search tree through four phases: \textit{Selection}, \textit{Expansion}, \textit{Simulation}, and \textit{Backpropagation}.

\paragraph{Selection.}
In each iteration, the algorithm traverses the tree from the root to a leaf node by recursively selecting the child node that maximizes the Upper Confidence Bound (UCB)~\citep{silver2017mastering}. Formally, for a parent node $s$ and a child node $s'$ (reached by action $a$), the selection policy is defined as:
\begin{equation}
    a^* = \operatorname*{argmax}_{a} \left( Q(s, a) + c_{\text{puct}} \cdot P(s, a) \frac{\sqrt{N(s)}}{1 + N(s, a)} \right),
\end{equation}
where $Q(s, a)$ is the estimated value of the action, $P(s, a)$ is the prior probability given by the LLM, $N(s)$ is the visit count of the parent node, and $c_{\text{puct}}$ is the exploration constant. This mechanism balances the exploitation of high-quality reasoning paths with the exploration of uncertain branches.

\paragraph{Expansion.}
Once a leaf node $s_{\text{leaf}}$ is reached, the Solver expands it by sampling $k$ potential next-step thoughts $\{\tau^{(1)}, \dots, \tau^{(k)}\}$ from the LLM policy $\pi_\theta(\cdot | s_{\text{leaf}})$. These thoughts typically represent intermediate reasoning steps or algorithm designs. Each new thought $\tau$ creates a new child node appended to the current trajectory.

\paragraph{Simulation and Code Synthesis.}
Standard MCTS typically performs random rollouts to estimate value. In our context, however, we leverage the LLM's completion capability to perform a \textit{semantic simulation}. For a newly expanded thought node $s_{\text{new}}$, the Solver performs a rollout to generate a complete code candidate $C$.
This design ensures that valid code candidates are generated continuously throughout the search process, rather than only at the search depth limit. 
Consequently, generated candidates are subjected to immediate screening: only those that pass both the public tests $\mathcal{T}_{\text{pub}}$ and previous accumulated adversarial suite in the \emph{Global Test Filter} $\mathcal{T}_{\text{global}}$ are deposited into the shared \emph{Code Pool} to support the Attacker's subsequent operations.

\paragraph{Backpropagation with Hybrid Feedback.}
After simulation, the generated code $C$ is evaluated to compute a reward $R$. The reward is a composite signal from two sources:
\begin{itemize}[leftmargin=*, topsep=0pt, partopsep=0pt, itemsep=0pt, parsep=4pt]
    \item \textbf{Intrinsic Correctness ($R_{\text{pub}}$):} The pass rate on the visible public test cases $\mathcal{T}_{\text{pub}}$.
    \item \textbf{Extrinsic Interaction ($V_{\text{penalty}}$):} A penalty signal fed back from the Global Test Filter (see Figure \ref{fig:overview}, blue arrow ``Backpropagation(-V)''). If the generated code $C$ is later found to fail on the global adversarial tests $\mathcal{T}_{\text{global}}$ or the newly generated $T_{\text{new}}$, a penalty $-V$ is applied to the current node of the solver.
\end{itemize}
The final value estimate $Q(s, a)$ of all nodes along the trajectory is updated using the collected reward, penalizing reasoning paths that lead to fragile code and reinforcing those that survive the adversarial filtering.

\subsection{Attacker MCTS: Vulnerability Discovery}
\label{subsec:attacker}

Parallel to the Solver, the Attacker operates to expose defects in the generated codes. Unlike standard test generation, which blindly targets the problem description, our Attacker employs a targeted adversarial strategy conditioned on the evolving \emph{Code Pool}.

\paragraph{Persistent Search with an Evolving Code Pool.}
The Attacker maintains a persistent search tree that grows incrementally throughout the searching process. The root state represents the problem context $S_{\text{test}} = \{P, \mathcal{C}_{\text{pool}}\}$, where $\mathcal{C}_{\text{pool}}$ is the dynamic set of currently accepted code that have passed all public tests. 
Crucially, this tree is \textit{retained} across iterations. When $\mathcal{C}_{\text{pool}}$ reaches a certain capacity, the Attacker resumes search from the existing tree structure, allowing it to progressively refine its attack strategies against an increasingly robust population of codes.

\paragraph{Selection and Expansion.}
The selection phase mirrors the Solver's logic, using a UCB-based policy to navigate to the most promising attacker thought nodes $\sigma$. 
At the frontier, the Attacker expands a new thought node $\sigma_{\text{new}}$ representing a specific testing strategy (e.g., ``Check boundary condition for $N=0$'' or ``Test with large prime inputs''). This reasoning step guides the subsequent generation towards specific vulnerability types rather than random fuzzing.

\paragraph{Simulation: Divergence-driven Test Synthesis.}
Upon reaching a new thought node $\sigma_{\text{new}}$, the Attacker employs a Divergence-driven Multi-sample Test Synthesis (DMTS) strategy to ensure high-quality test generation. The LLM generates a batch of $k$ candidate test inputs $\{T^{(1)}, \dots, T^{(k)}\}$ conditioned on the current code pool: $ \{T^{(1)}, \dots, T^{(k)}\} \sim \pi_{\text{adv}}(\cdot | P, \mathcal{C}_{\text{pool}}, \sigma_{\text{new}})$.

By explicitly conditioning on $\mathcal{C}_{\text{pool}}$, the Attacker aims to craft inputs that exploit the logical discrepancies observed among the current code. Generating multiple candidates reduces the variance of the generation process and increases the probability of discovering a valid corner case.

Since the code pool consists of public-test–passing candidates and no ground-truth output is available at test time, we reward inputs that expose pseudo-correctness by causing the candidates in the pool to produce diverging outputs. With this principle, we propose to evaluate the \emph{Divergence Reward} $R_{div}$ for each candidate. The reward is defined based on the disagreement of outputs produced by the Code Pool:
\begin{equation}
    R_{\text{div}}(T) = \mathbb{I}\left[ \exists i, j : O_i(T) \neq O_j(T) \right] \cdot w_{\text{div}},
\end{equation}
where $O_i(T)$ denotes the execution output of code $C_i \in \mathcal{C}_{\text{pool}}$ on input $T$; $w_{\text{div}}$ is to normalize this reward by the number of unique outputs. 
We select the most discriminative test case $T^* = \operatorname*{argmax}_{\text{m}} R_{\text{div}}(T^{(\text{m})})$ from the batch.

While output divergence cannot detect the case where all candidates agree on the same wrong output, it is aligned with our test-time objective: distinguishing pseudo-correct code that already pass $\mathcal{T}_{\text{pub}}$. In practice, disagreement provides a high-signal trigger for discovering latent logical gaps, and we further guard against invalid via the Arbiter-based validity check.

\paragraph{Backpropagation.}
 The reward $R_{\text{div}}(T^*)$ is backpropagated up the Attacker's tree, reinforcing the reasoning strategies that lead to high-divergence scenarios.

\subsection{Interaction and Global Filtering Hub}
\label{subsec:interaction}

The interaction module acts as the central ``arena'' where the Solver and Attacker continuously exchange feedback. This mechanism ensures that only valid, high-quality test cases are retained to penalize fragile code.

\paragraph{LLM-based Output Arbiter.}
When the Attacker generates a test case $T_{\text{new}}$ that induces output divergence among the code candidates (i.e., $\exists C_i, C_j: O_i(T_{\text{new}}) \neq O_j(T_{\text{new}})$), a ground truth label is required to determine which candidate is at fault. Since hidden tests are unavailable during inference, we introduce an \emph{LLM-based Output Arbiter} to adjudicate the results.
The Arbiter takes the problem description $P$, the generated input $T_{\text{new}}$, and the divergent outputs $\{O_i, O_j, \dots\}$ as input. It analyzes the semantic logic of $P$ to identify the correct expected output $O^*$.
\begin{equation}
    O^*, \text{validity} \leftarrow \text{Arbiter}(P, T_{\text{new}}, \{O_i, O_j\}).
\end{equation}

If the Arbiter deems the test input $T_{\text{new}}$ ambiguous or invalid, the test is discarded. Otherwise, the pair $(T_{\text{new}}, O^*)$ is formalized as a new valid test case.

\paragraph{Global Filtering and Feedback Loops.}
Successfully adjudicated tests are added to the \emph{Global Test Filter} ($\mathcal{T}_{\text{global}}$), a dynamic repository of ``hard'' corner cases found during the search.
\begin{itemize}[leftmargin=*, topsep=0pt, partopsep=0pt, itemsep=0pt, parsep=4pt]
    \item \textbf{Penalty Feedback (-V) for Solver:} With the adjudicated ground truth $O^*$, the system identifies the specific candidates $\mathcal{C}_{\text{fail}}$ that produced incorrect outputs. A penalty value $-V$ is backpropagated to their corresponding thought nodes in the Solver's MCTS tree. This signal discourages the Solver from pursuing reasoning paths that led to these fragile implementations.
    \item \textbf{Filtering Mechanism:} The $\mathcal{T}_{\text{global}}$ acts as a gatekeeper. In subsequent iterations, any new candidate generated by the Solver must pass all tests in $\mathcal{T}_{\text{global}}$ before entering the Code Pool. This ensures that the Code Pool monotonically improves in robustness.
\end{itemize}

\subsection{Inference: Robust Selection}
\label{subsec:inference}

After the search process, the Solver yields a diverse set of candidate programs. To select the final submission $C^*$, we employ a hierarchical \emph{Test-based Reranking} strategy that prioritizes robustness against the accumulated adversarial knowledge.

The code candidates are ranked by a two-stage criterion:
\begin{itemize}[leftmargin=*, topsep=0pt, partopsep=0pt, itemsep=0pt, parsep=4pt]
    \item \textbf{Primary Sort (Public Integrity):} Candidates are first ranked by their pass rate on the public test suite $\mathcal{T}_{\text{pub}}$. This step filters out solutions that fail to meet the basic problem requirements.
    \item \textbf{Secondary Sort (Adversarial Robustness):} For candidates with identical public test scores (which is common due to the scarcity of $\mathcal{T}_{\text{pub}}$), we further rank them by their pass rate on the generated adversarial tests in $\mathcal{T}_{\text{global}}$.
\end{itemize}

This mechanism effectively resolves ties among ``pseudo-correct'' solutions that overfit the public tests, selecting the candidate that survives the hostile environment constructed by the Attacker.

\section{Experiments}
\label{sec:experiments}

\subsection{Experiment Settings}

\paragraph{Datasets.}
We evaluate \our{} on two challenging competition-level code generation benchmarks: APPS \citep{hendrycks2021measuring} and TACO \citep{li2023taco}. The APPS dataset contains three levels of difficulties: \textit{Introductory}, \textit{Interview}, and \textit{Competition}. And \textit{Easy}, \textit{Medium}, and \textit{Hard} split for TACO. Followin prior work~\citep{li2025rethinkmcts}, we evaluate all the methods on the formal 100 problems per split. For each problem, we set the maximum number of $|\mathcal{T}_{\text{pub}}|=5$.
Following prior work~\citep{austin2021program, chen2021evaluating, dong2025codescore}, we use the standard \textit{pass rate} and \textit{pass@1} metrics evaluated on the hidden test suite to measure robust correctness.

\begin{table*}[t]
\centering
\caption{Main Results. We report Pass Rate and Pass@1 accuracy using Qwen3-4B-Instruct/8B backbones under rollout 16. \our{} consistently outperforms state-of-the-art search baselines, showing significant gains on the most challenging subsets.}
\resizebox{\linewidth}{!}
{%
\begin{tabular}{l|cccc|cccc|cccc|cccc}
\toprule[1.5pt]
\multicolumn{1}{c|}{\multirow{3}[4]{*}{\textbf{Model}}} & \multicolumn{8}{c|}{\textbf{APPS}} &\multicolumn{8}{c}{\textbf{TACO}} \\
\multicolumn{1}{l|}{} & \multicolumn{4}{c}{\textbf{Pass Rate (\%)}} & \multicolumn{4}{c|}{\textbf{Pass@1 (\%)}} &\multicolumn{4}{c}{\textbf{Pass Rate (\%)}} & \multicolumn{4}{c}{\textbf{Pass@1 (\%)}} \\
\cline{2-17}
\multicolumn{1}{l|}{} & \multicolumn{1}{c}{Intro.}  & \multicolumn{1}{c}{Inter.} & \multicolumn{1}{c}{Comp.} & \multicolumn{1}{c|}{Avg.} & \multicolumn{1}{c}{Intro.} & \multicolumn{1}{c}{Inter.} & \multicolumn{1}{c}{Comp.} & \multicolumn{1}{c|}{Avg.} & \multicolumn{1}{c}{Easy} &\multicolumn{1}{c}{Medium} & \multicolumn{1}{c}{Hard}  & \multicolumn{1}{c|}{Avg.} & \multicolumn{1}{c}{Easy} &\multicolumn{1}{c}{Medium} & \multicolumn{1}{c}{Hard} & \multicolumn{1}{c}{Avg.}\\
\hline
\multicolumn{1}{l|}{\textbf{Qwen3-4B-Instruct}} &\multicolumn{4}{l|}{} &\multicolumn{4}{l|}{}  &\multicolumn{4}{l|}{} &\multicolumn{4}{l}{}\\
\hline
Base  & 51.94 & 53.30 & 31.70 & 45.64 & 35 & 27 & 15 & 25.67 & 57.27  & 39.97  & 29.87  & 42.37 & 33 & 20 & 10 & 21.00\\
Base (16)  & 65.27 & 66.52 & 48.77 & 60.18 & 46 & 40 & 23 & 36.33 & 68.65  & 56.15  & 48.49 & 57.76 & 37 & 24 & 12 & 24.33\\
PG-TD  & 71.88 & 69.05 &40.50 & 60.48 & 56 & 43 & 27 & 42.00 & 70.99  & 61.81  & 56.54 & 63.11 & 50 & 35 & 20 & 35.00\\
MCTS-Thought  & 75.56 & 74.37 & 58.10 & 69.34 & 57 & 44 & 29 & 43.44 & 82.13  & 68.10  & 65.83  & 72.02 & 47 & 32 & 14 & 31.00\\
RethinkMCTS & 77.77 & 72.69 & 61.13 & 70.53 & 56 & 47 & 29 & 43.67 & 83.91  & 68.79  & 64.75 & 72.48 & 52 & 31 & 18 & 33.67\\
ToT  & 65.63 & 66.46 & 46.83 & 59.64 & 46 & 43 & 33 & 40.67 & 77.47  & 64.43  & 56.98 & 66.29 & 52 & 34 & 14 & 34.00\\
LATS  & 71.88 & 68.58 & 53.25 & 64.57 & 50 & 39 & 27 & 38.67 & 75.46  & 63.94  & 58.34 & 65.91 & 44 & 29 & 17 & 30.00\\
\rowcolor{Gray} 
\our{}  & \textbf{78.63} & \textbf{75.24} & \textbf{64.45} & \textbf{72.77} & \textbf{61} & \textbf{52} & \textbf{36} & \textbf{49.67} & \textbf{84.04}  & \textbf{73.05}  & \textbf{66.23} & \textbf{74.44} & \textbf{55} & \textbf{37} & \textbf{22} & \textbf{38.00}\\
\hline
\multicolumn{1}{l|}{\textbf{Qwen3-8B}} &\multicolumn{4}{l|}{} &\multicolumn{4}{l|}{}  &\multicolumn{4}{l|}{} &\multicolumn{4}{l}{}\\
\hline
Base  & 42.97 & 40.21 & 25.95 & 36.38 & 21 & 17 & 7 & 15.00 & 45.06  & 29.08  & 24.97 & 33.04 & 22 & 9 & 4 & 11.67\\
Base (16)  & 50.18 & 46.75 & 38.87 & 45.27 & 26 & 21 & 15 & 20.67 & 51.13  & 35.19  & 34.97 & 40.43 & 23 & 8 & 7 & 12.67\\
PG-TD  & 56.08 & 53.81 & 32.67 & 47.52 & 30 & 29 & 14  & 24.33 & 63.59  & 46.11  & 44.09 & 44.09 & 39 & 18 & 13 & 23.33\\
MCTS-Thought  & 62.80 & 60.55 & 50.40 & 57.92 & 37 & 36 & 21 & 31.33 & 70.15  & 56.11  & 49.69 & 58.65 & 36 & 21 & 8 & 21.67\\
RethinkMCTS  & 66.68 & 62.16 & 52.18 & 60.34 & 37 & 35 & 23 & 31.67 & 76.35  & 60.82  & 48.89 & 62.02 & 42 & 22 & 12 & 25.33\\
ToT  & 50.52 & 49.22 & 33.67 & 44.47 & 28 & 22 & 16 & 22.00 & 66.49  & 46.73  & 43.43 & 52.21 & 39 & 21 & 13 & 24.33\\
LATS  & 59.64 & 56.01 & 42.30 & 52.65 & 30 & 31 & 17 & 26.00 & 62.52  & 51.08  & 38.03 & 50.54 & 32 & 21 & 7 & 20.00\\
\rowcolor{Gray} 
\our{}  & \textbf{66.96} & \textbf{63.15} & \textbf{53.27} & \textbf{61.13} & \textbf{44} & \textbf{40} & \textbf{24} & \textbf{36.00} & \textbf{76.99}  & \textbf{60.84}  & \textbf{52.95} & \textbf{63.94} & \textbf{44} & \textbf{27} & \textbf{13} & \textbf{28.00}\\
\bottomrule[1.5pt]
\end{tabular}%
}
\vspace{-5mm}
\label{tab:main_results}
\end{table*}

\paragraph{Baselines.}
We compare \our{} against state-of-the-art methods spanning three categories:
(1) Direct Synthesis: Standard \textit{Zero-shot} prompting and \textit{Best-of-N} sampling (with $N=16$) to establish performance lower bounds.
(2) Search and Planning: Advanced search-based frameworks including PG-TD \citep{zhang2023planning}, Tree of Thoughts (ToT) \citep{yao2023tree}, and LATS \citep{zhou2023language}, which utilize lookahead planning or self-reflection.
(3) MCTS Variants: We also compare with MCTS-Thought (MCTS for reasoning search) and RethinkMCTS \citep{li2025rethinkmcts}, a recent method focusing on repairing erroneous nodes.
Due to space constraints, detailed descriptions of these baselines and their specific configurations are provided in Appendix~\ref{app:baselines}.

\paragraph{Implementation Details.}
We employ the Qwen3-4B-Instruct-2507 and Qwen3-8B (non-instruct)~\citep{yang2025qwen3} as the main backbone LLM for both the Solver and Attacker agents. We also experiment on DeepSeek-V3.2~\cite{liu2025deepseek} in the scaling experiment. 
For the Solver, we implement a standard MCTS with a UCB exploration constant $c_{puct}=4$. 
For the Attacker, we set a moderate search budget of $N=2$ rollouts per iteration, balancing computational efficiency with adversarial strength.
The maximum interaction depth is set to 16 rollouts and penalty value $V=0.1$.
All experiments are conducted using the vLLM library~\citep{kwon2025vllm} for efficient inference.

\subsection{Main Results}
Table ~\ref{tab:main_results} reports results on APPS and TACO with two backbones. Overall, \our{} achieves the best performance across benchmarks and backbones, with especially strong gains on APPS and on the TACO average. Comparing baselines, we observe a clear granularity effect: token-level lookahead (PG-TD) offers limited gains due to its myopic horizon; code-level search (LATS) improves performance but struggles with the sparsity of the full program space; whereas thought-level search (MCTS-Thought) proves most effective by decomposing complex logic into manageable planning steps. Crucially, unlike these methods that primarily allocate compute to expanding candidate volume, \our{} strategically directs budget toward \textit{adversarial verification}. This active purification mechanism filters out pseudo-correct solutions, enabling a significantly higher conversion of public-test pass rates into robust hidden-test correctness (Pass@1).

\subsection{Scaling with Backbone Capability}
\begin{figure}[t]
    \centering
    \includegraphics[width=0.49\textwidth]{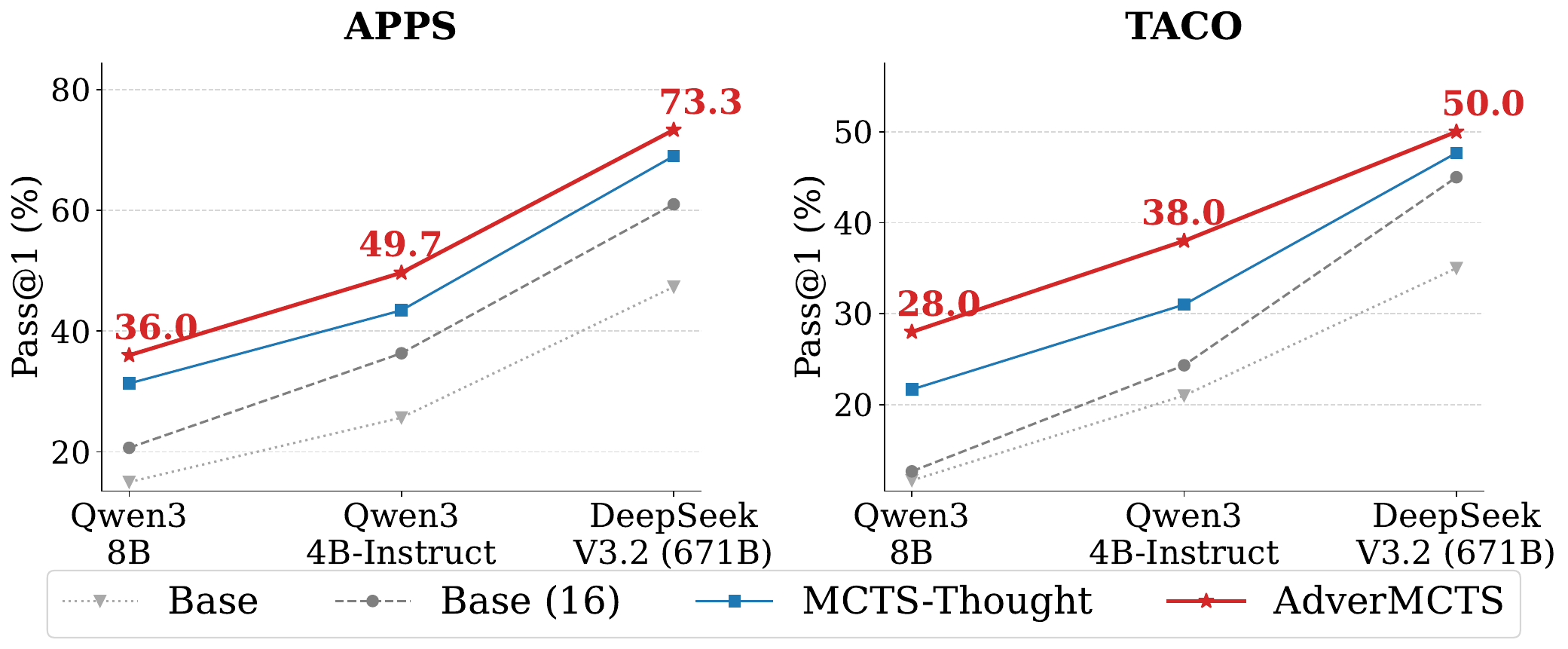}
    \caption{Scalability with Model Capabilities. We compare methods across backbones sorted by intrinsic capability. Despite parameter discrepancies, \our{} consistently amplifies performance, maintaining a significant lead across all model scales.
    }
    \vspace{-7mm}
    \label{fig:scaling}
\end{figure}

To verify the scalability and model-agnosticism of \our{}, we extended our evaluation to the state-of-the-art frontier model, DeepSeek-V3.2 (671B)~\citep{liu2025deepseek}. As shown in Figure~\ref{fig:scaling}, we order the models on the x-axis by their baseline sampling performance to visualize the correlation between intrinsic model capability and method gain. The results reveal a strictly monotonic upward trend: \our{} consistently outperforms both the strong sampling baseline (Base-16) and the search baseline (MCTS-Thought) across all settings. Notably, the performance gain remains robust even on the 671B frontier model, confirming that our adversarial mechanism is agnostic to the underlying model capacity and effectively scales from smaller open weights to massive state-of-the-art base models.

\begin{figure*}[t]
    \centering
    \includegraphics[width=0.9\textwidth]{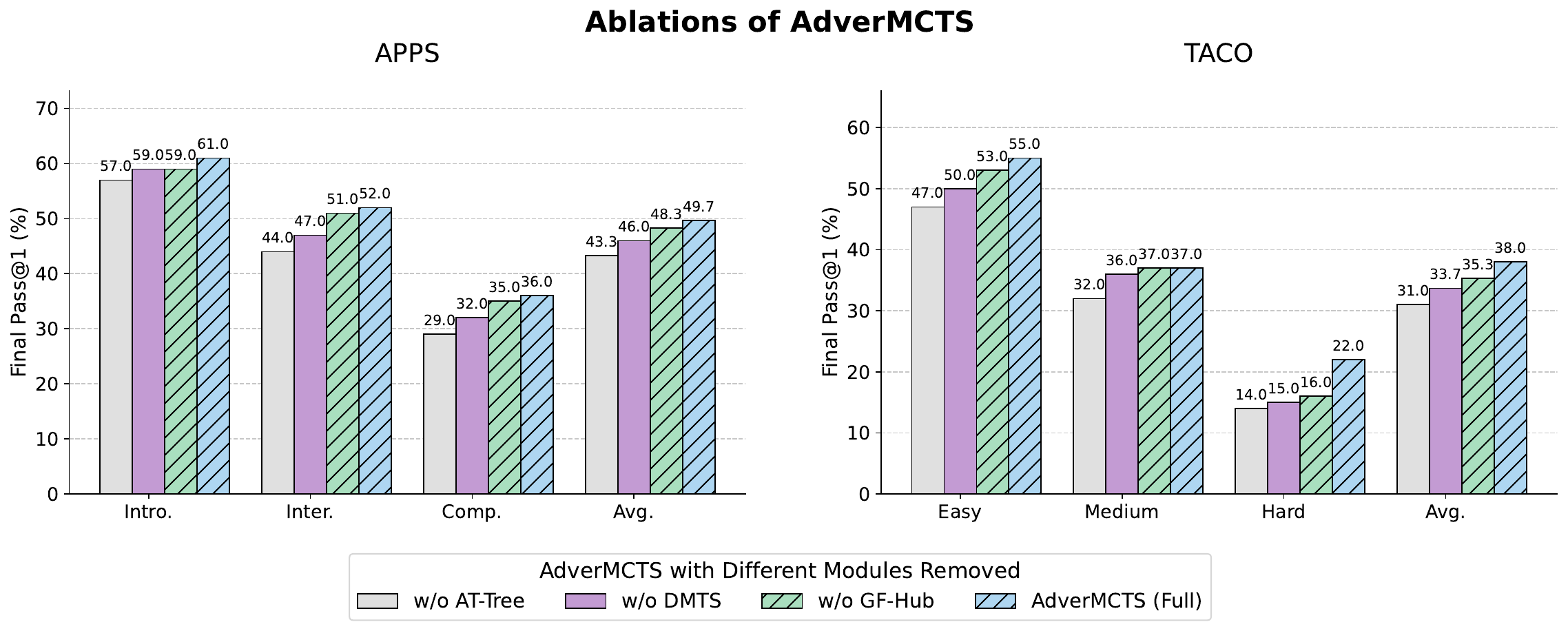}
    \caption{Ablation study of \our{} on APPS and TACO. We report the final Pass@1 (\%) under the same inference budget while removing one component at a time.}
    \vspace{-5mm}
    \label{fig:ablation}
\end{figure*}
\subsection{Ablation Study}
\label{subsec:ablation}
We ablate three core components of \our{}, removing exactly one module at a time while keeping all other settings fixed: (i) \textit{AT-Tree}, which removes the entire Attacker agent and adversarial testing, reverting to standard search guided only by public tests; (ii) \textit{DMTS}, which disables the multi-sample test synthesis, relying on a single stochastic sample for test generation instead of divergence-driven selection; and (iii) \textit{GF-Hub}, where we specifically disable \emph{pre-admission} global screening before a candidate enters the code pool, while keeping the final test-based re-ranking unchanged.

As shown in Figure~\ref{fig:ablation}, all modules contribute positively. Removing \textit{AT-Tree} yields the largest drop, since no adversarial tests are produced and the solver degenerates to verification on public tests only, making pseudo-correct candidates much harder to eliminate. Disabling \textit{DMTS} also consistently hurts performance, indicating that multi-sample synthesis is important to reduce variance and reliably surface divergence-triggering inputs. Finally, turning off the \textit{GF-Hub} screening step further degrades results, especially on harder splits, suggesting that early global screening helps prevent fragile solutions from polluting the pool and weakening subsequent adversarial refinement and selection.

\subsection{Extended Analysis}
\begin{figure}[htbp]
    \centering
    \includegraphics[width=0.49\textwidth]{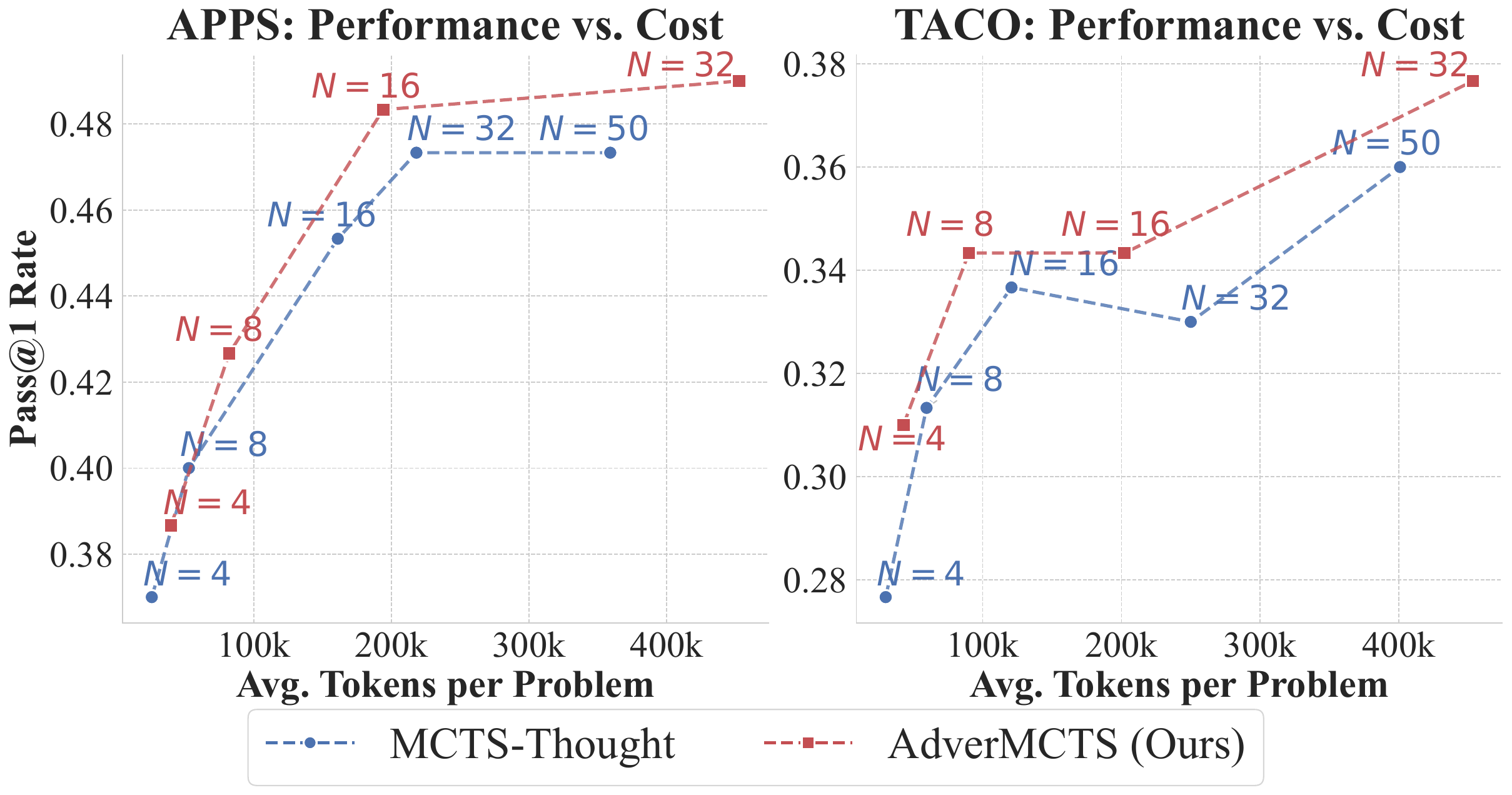}
    \caption{Cost-Performance Pareto Frontiers (upper left is better). Comparison of Pass@1 accuracy against average token consumption per problem. \our{} consistently achieves a superior Pareto frontier. The labels denote the rollout budget.
    }
    \vspace{-10pt}
    \label{fig:cost}
\end{figure}

\paragraph{Cost-Performance Trade-off under Test-Time Compute Scaling.}
We investigate the scaling efficiency of \our{} by analyzing the Pareto frontier~\citep{lotov2008visualizing} between solution correctness (Pass@1) and computational cost (average token consumption). As shown in Figure~\ref{fig:cost}, across both APPS and TACO, \our{} consistently dominates the baseline frontier: for a similar token budget, it achieves higher Pass@1, indicating that the extra compute is spent on more effective verification rather than producing more redundant samples. Notably, on both datasets, \our{} with a budget of $N=16$ rollouts achieves superior accuracy compared to the baseline with $N=32$ rollouts, while consuming significantly fewer tokens. This observation creates a compelling counter-argument to the concern of overhead introduced by the dual-agent architecture: the active ``purification'' of the search space via adversarial counter-examples proves to be more compute-efficient than brute-force scaling of reasoning paths. By filtering out pseudo-correct solutions early, \our{} achieves ``smarter'' test-time scaling rather than simply ``harder'' scaling.

\paragraph{Co-evolution of Test and Code.}
To uncover the temporal mechanism behind \our{}, we visualize the iterative interaction between the Attacker and the Solver. We selected 100 distinct problems from the hardest subsets (APPS-Competition and TACO-Hard) and tracked two key metrics over 16 rollouts: 
1) The Cumulative Number of Valid Adversarial Test Corner Cases (that induce divergence among codes) discovered by the Attacker (Orange); and 
2) The Hidden Pass Rate of the Solver's code pool (Blue), which serves as a proxy for true semantic correctness.

\begin{figure}[htbp]
    \centering
    \includegraphics[width=0.49\textwidth]{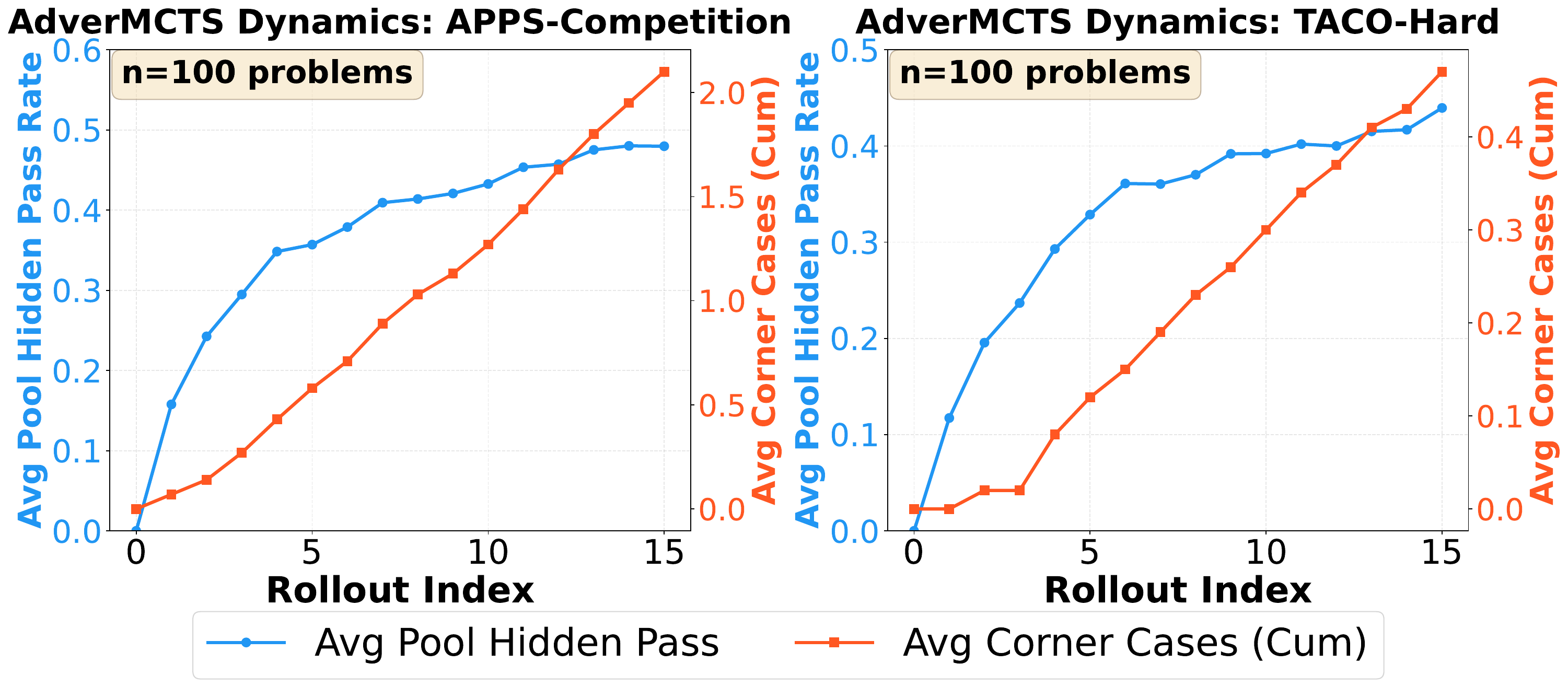}
    \caption{Visualization of Adversarial Dynamics. We track the co-evolution process. The synchronous rise of cumulative corner cases (\textcolor[HTML]{ed7d31}{\textbf{Orange}}) and the Solver's hidden pass rate (\textcolor[HTML]{4472c4}{\textbf{Blue}}) demonstrates that accumulating adversarial constraints drives the Solver toward robust solutions.
    }
    \vspace{-10pt}
    \label{fig:dynamics}
\end{figure}

Figure~\ref{fig:dynamics} presents the dual-axis trajectories. We observe a strong positive correlation between the accumulation of adversarial constraints and the improvement of solution robustness. 
Initially (Rollout 0-3), the code pool is fragile, passing a few hidden tests. As the Attacker actively identifies valid corner cases (indicated by the steep rise in the orange curve), the filtering mechanism filters out pseudo-correct codes. 
By Rollout 16, the accumulation of diverse test cases acts as a comprehensive logical boundary, guiding the Solver's pass rate to converge at a significantly higher level. This confirms that \our{} functions as a constructive adversarial process: adversarial test cases evolve into strict constraints, effectively pruning fragile solutions and forcing the Solver to generalize beyond the public tests.

\begin{figure}[h]
    \centering
    \includegraphics[width=0.42\textwidth]{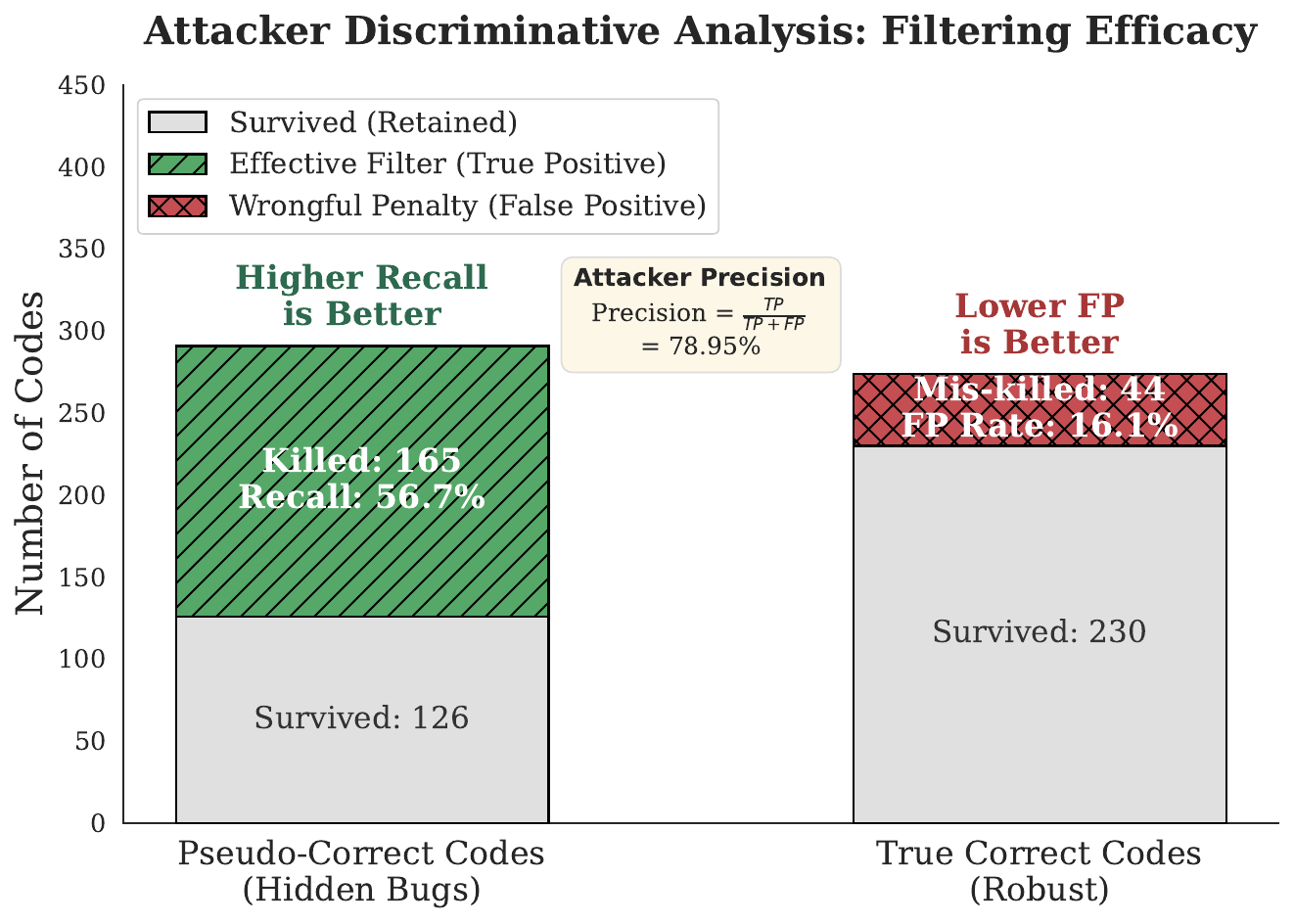}
    \caption{Attacker Discriminative Analysis. We evaluate filtering efficacy on \textit{Pseudo-Correct} (pass public, fail hidden) versus \textit{True Correct} codes. The \textcolor[HTML]{2d6a4f}{\textbf{green}} hatched area marks effective bug identification (True Positives), while the \textcolor[HTML]{a63737}{\textbf{red}} area indicates wrongful penalization (False Positives).}
    \label{fig:attacker_analysis}
    \vspace{-10pt}
\end{figure}

\textbf{Discriminative Power of Adversarial Test Generation.} 
To show that the Attacker provides a \emph{meaningful} supervision signal (rather than stochastic noise), we evaluate how well its generated adversarial tests discriminate \textit{Pseudo-Correct} solutions (pass $\mathcal{T}_{\text{pub}}$ but fail $\mathcal{T}_{\text{hidden}}$) from \textit{True Correct} ones (pass all tests). Figure~\ref{fig:attacker_analysis} visualizes the impact of the generated adversarial tests on these two groups. The results highlight three key observations: (1) \textbf{High Sensitivity:} It achieves a Recall of 56.7\% (165/291), exposing over half of the ``silent bugs'' missed by standard MCTS; (2) \textbf{Conservative Behavior:} It maintains a low False Positive Rate of 16.1\%, indicating the generated corner cases are largely valid; and (3) \textbf{High Precision:} The overall precision of 78.95\% confirms that output divergence is a reliable proxy for semantic correctness without ground truth.

\paragraph{Solver-Aware is Crucial for Effective Attacker.}
We investigate two critical design choices to see the impact of solver awareness for the attacker: (1) Solver Context, by ablating the Attacker's access to the code pool $\mathcal{C}_{\text{pool}}$ (``w/o Solver Context''); and (2) Search Algorithm, we examine whether this solver-to-attacker signal can be \emph{effectively exploited} by different search strategies, by replacing MCTS with lighter Best-of-N baseline and Random-MCTS baseline where the \emph{UCB-based selection} in MCTS is replaced by random selection. 

Table~\ref{tab:attacker_ablation} yields two key insights. First, removing solver context significantly degrades performance. Without observing $\mathcal{C}_{\text{pool}}$, the Attacker generates generic test cases rather than targeted counter-examples that exploit the Solver's specific logical gaps. Second, MCTS outperforms Best-of-N and Random-MCTS not simply by searching ``harder'', but because its selection, value estimation, and backpropagation can absorb solver-derived feedback. This allows the Attacker to refine test reasoning progressively, whereas one-shot sampling is limited to the immediate quality of samples and cannot reliably leverage the feedback.

\begin{table}[t]
    \centering
    \caption{Ablation on Attacker Optimization. We analyze the impact of \textit{Solver Context} and \textit{Search Strategy}. The results validate that both contextual targeting and iterative tree search are essential for finding meaningful vulnerabilities.}
    \label{tab:attacker_ablation}
    \resizebox{0.48\textwidth}{!}{
    \begin{tabular}{l|ccc|c|ccc|c}
        \toprule
        \multirow{2}{*}{\textbf{Method}} & \multicolumn{4}{c|}{\textbf{APPS (Pass@1)}} & \multicolumn{4}{c}{\textbf{TACO (Pass@1)}} \\
        \cmidrule(lr){2-5} \cmidrule(lr){6-9}
         & \textbf{Intro.} & \textbf{Inter.} & \textbf{Comp.} & \textbf{Avg.} & \textbf{Easy} & \textbf{Med.} & \textbf{Hard} & \textbf{Avg.} \\
        \midrule
        \textbf{AdverMCTS (Full)} & \textbf{61.0} & \textbf{52.0} & \textbf{36.0} & \textbf{49.7} & \textbf{55.0} & \textbf{37.0} & \textbf{22.0} & \textbf{38.0} \\
        \midrule
        \multicolumn{9}{l}{\textit{Information Cutoff}} \\
        w/o Solver Context & 58.0 & 49.0 & 31.0 & 46.0 & 52.0 & 35.0 & 17.0 & 34.7 \\
        \midrule
        \multicolumn{9}{l}{\textit{Search Strategy}} \\
        Best-of-N & 58.0 & 48.0 & 31.0 & 45.7 & 53.0 & 31.0 & 18.0 & 34.0 \\
        Random-MCTS & 57.0 & 47.0 & 27.0 & 43.7 & 47.0 & 34.0 & 16.0 & 32.3 \\
        \bottomrule
    \end{tabular}
    }
    \vspace{-4mm}
\end{table}




\paragraph{Output Judging Matters: LLM Arbiter vs. Majority Voting.}
In \our{}, accurately determining the expected output for a generated test input is critical. Once a test input induces disagreement among codes in the pool, we must determine which output is correct to label the test and update the global test filter. We compare two judging strategies: (\textit{i}) \textit{Majority Voting}, which selects the majority output among the current code pool, and (\textit{ii}) our \textit{LLM Arbiter}, where the same backbone LLM adjudicates the correct output by reading the problem specification and the candidate outputs.
As shown in Table~\ref{tab:arbiter_ablation}, the Arbiter significantly outperforms Voting on both benchmarks. 
Crucially, on the TACO dataset, we analyzed the validity of the test pairs generated by both methods. The Voting strategy resulted in a low rate of valid test pairs, indicating that the majority of solutions frequently converged on incorrect outputs. These results justify the design choice of using an LLM arbiter: it improves both the quality of adversarial labels and the effectiveness of adversarial verification.

\begin{table}[htbp]
    \centering
    \caption{Comparing output-judging strategies for adversarial tests. The LLM arbiter yields higher downstream performance and substantially higher labeled-test validity(measured against gold code).}
    \label{tab:arbiter_ablation}
    \resizebox{\columnwidth}{!}{%
    \begin{tabular}{l|c|cc}
        \toprule
        \multirow{2}{*}{\textbf{Oracle Strategy}} & \textbf{Signal Quality (TACO)} & \multicolumn{2}{c}{\textbf{Performance (Pass@1)}} \\
        \cmidrule(lr){2-2} \cmidrule(lr){3-4}
         & Labeled-test validity ($\uparrow$) & APPS Avg. & TACO Avg. \\
        \midrule
        Majority Voting & 37.88\% & 43.67 & 32.67 \\
        \textbf{LLM Arbiter (Ours)} & \textbf{79.88\%} & \textbf{49.67} & \textbf{37.33} \\
        \bottomrule
    \end{tabular}%
    }
    \vspace{-5mm} 
\end{table}

\section{Conclusion}
In this paper, we presented \our{}, a novel framework that addresses pseudo-correctness in code generation by shifting the paradigm from static verification to adversarial purification. 
By coupling a Solver MCTS with an active Attacker MCTS, \our{} constructs a progressively hostile environment that co-evolves with the solution candidates.
This minimax interaction allows the system to autonomously discover corner cases and mitigate survivorship bias without human-annotated test oracles.
Extensive experiments demonstrate that \our{} outperforms state-of-the-art search-based methods, validating the effectiveness of adversarial test-time compute.
We believe this work opens new avenues for enhancing LLM reasoning robustness through autonomous, adversarial self-improvement.

\section*{Impact Statement}
This paper presents a method for improving the robustness and correctness of automated code generation systems. 
Our work contributes to the reliability of AI-assisted programming, potentially reducing software bugs and development costs.
However, as with any advanced code generation technology, there is a potential risk of misuse for generating malicious software.
We believe the benefits of robust verification outweigh these risks, as our ``Attacker'' agent focuses on logical correctness rather than security exploitation.
There are no specific societal consequences that we feel must be highlighted here.






\bibliography{example_paper}

\begin{thebibliography}{60}
\providecommand{\natexlab}[1]{#1}
\providecommand{\url}[1]{\texttt{#1}}
\expandafter\ifx\csname urlstyle\endcsname\relax
  \providecommand{\doi}[1]{doi: #1}\else
  \providecommand{\doi}{doi: \begingroup \urlstyle{rm}\Url}\fi

\bibitem[Austin et~al.(2021)Austin, Odena, Nye, Bosma, Michalewski, Dohan,
  Jiang, Cai, Terry, Le, et~al.]{austin2021program}
Austin, J., Odena, A., Nye, M., Bosma, M., Michalewski, H., Dohan, D., Jiang,
  E., Cai, C., Terry, M., Le, Q., et~al.
\newblock Program synthesis with large language models.
\newblock \emph{arXiv preprint arXiv:2108.07732}, 2021.

\bibitem[Brown et~al.(2024)Brown, Juravsky, Ehrlich, Clark, Le, R{\'e}, and
  Mirhoseini]{brown2024large}
Brown, B., Juravsky, J., Ehrlich, R., Clark, R., Le, Q.~V., R{\'e}, C., and
  Mirhoseini, A.
\newblock Large language monkeys: Scaling inference compute with repeated
  sampling.
\newblock \emph{arXiv preprint arXiv:2407.21787}, 2024.

\bibitem[Browne et~al.(2012)Browne, Powley, Whitehouse, Lucas, Cowling,
  Rohlfshagen, Tavener, Perez, Samothrakis, and Colton]{browne2012survey}
Browne, C.~B., Powley, E., Whitehouse, D., Lucas, S.~M., Cowling, P.~I.,
  Rohlfshagen, P., Tavener, S., Perez, D., Samothrakis, S., and Colton, S.
\newblock A survey of monte carlo tree search methods.
\newblock \emph{IEEE Transactions on Computational Intelligence and AI in
  games}, 4\penalty0 (1):\penalty0 1--43, 2012.

\bibitem[Chaffin et~al.(2022)Chaffin, Claveau, and Kijak]{chaffin2022ppl}
Chaffin, A., Claveau, V., and Kijak, E.
\newblock Ppl-mcts: Constrained textual generation through discriminator-guided
  mcts decoding.
\newblock In \emph{Proceedings of the 2022 {C}onference of the North American
  chapter of the {A}ssociation for {C}omputational {L}inguistics: {H}uman
  {L}anguage {T}echnologies}, pp.\  2953--2967, 2022.

\bibitem[Chen et~al.(2022)Chen, Zhang, Nguyen, Zan, Lin, Lou, and
  Chen]{chen2022codet}
Chen, B., Zhang, F., Nguyen, A., Zan, D., Lin, Z., Lou, J.-G., and Chen, W.
\newblock Codet: Code generation with generated tests.
\newblock \emph{arXiv preprint arXiv:2207.10397}, 2022.

\bibitem[Chen et~al.(2024{\natexlab{a}})Chen, Hu, Li, Gao, Xia, and
  Lo]{chen2024code}
Chen, J., Hu, X., Li, Z., Gao, C., Xia, X., and Lo, D.
\newblock Code search is all you need? {I}mproving code suggestions with code
  search.
\newblock In \emph{Proceedings of the IEEE/ACM 46th {I}nternational
  {C}onference on {S}oftware {E}ngineering}, pp.\  1--13, 2024{\natexlab{a}}.

\bibitem[Chen(2021)]{chen2021evaluating}
Chen, M.
\newblock Evaluating large language models trained on code.
\newblock \emph{arXiv preprint arXiv:2107.03374}, 2021.

\bibitem[Chen et~al.(2024{\natexlab{b}})Chen, Wang, Cao, Liu, Gao, Cui, Zhu,
  Ye, Tian, Liu, et~al.]{chen2024expanding}
Chen, Z., Wang, W., Cao, Y., Liu, Y., Gao, Z., Cui, E., Zhu, J., Ye, S., Tian,
  H., Liu, Z., et~al.
\newblock Expanding performance boundaries of open-source multimodal models
  with model, data, and test-time scaling.
\newblock \emph{arXiv preprint arXiv:2412.05271}, 2024{\natexlab{b}}.

\bibitem[Chen et~al.(2024{\natexlab{c}})Chen, White, Mooney, Payani, Su, and
  Sun]{chen2024tree}
Chen, Z., White, M., Mooney, R., Payani, A., Su, Y., and Sun, H.
\newblock When is tree search useful for {LLM} planning? it depends on the
  discriminator.
\newblock In \emph{Proceedings of the 62nd Annual Meeting of the Association
  for Computational Linguistics (Volume 1: Long Papers)}, pp.\  13659--13678,
  2024{\natexlab{c}}.

\bibitem[Dong et~al.(2025{\natexlab{a}})Dong, Ding, Jiang, Li, Li, and
  Jin]{dong2025codescore}
Dong, Y., Ding, J., Jiang, X., Li, G., Li, Z., and Jin, Z.
\newblock Codescore: Evaluating code generation by learning code execution.
\newblock \emph{ACM Transactions on Software Engineering and Methodology},
  34\penalty0 (3):\penalty0 1--22, 2025{\natexlab{a}}.

\bibitem[Dong et~al.(2025{\natexlab{b}})Dong, Jiang, Qian, Wang, Zhang, Jin,
  and Li]{dong2025survey}
Dong, Y., Jiang, X., Qian, J., Wang, T., Zhang, K., Jin, Z., and Li, G.
\newblock A survey on code generation with {LLM}-based agents.
\newblock \emph{arXiv preprint arXiv:2508.00083}, 2025{\natexlab{b}}.

\bibitem[Gao et~al.(2024)Gao, Gao, Gu, and Lyu]{gao2024search}
Gao, S., Gao, C., Gu, W., and Lyu, M.
\newblock Search-based {LLMs} for code optimization.
\newblock \emph{arXiv preprint arXiv:2408.12159}, 2024.

\bibitem[Gou et~al.(2024)Gou, Dong, Wu, and Ke]{gou2024rrgcode}
Gou, Q., Dong, Y., Wu, Y., and Ke, Q.
\newblock Rrgcode: Deep hierarchical search-based code generation.
\newblock \emph{Journal of Systems and Software}, 211:\penalty0 111982, 2024.

\bibitem[Han et~al.(2024)Han, Zhang, Meng, Zhang, Hu, and Weng]{han2024value}
Han, Y., Zhang, L., Meng, D., Zhang, Z., Hu, X., and Weng, S.
\newblock A value based parallel update mcts method for multi-agent cooperative
  decision making of connected and automated vehicles.
\newblock \emph{arXiv preprint arXiv:2409.13783}, 2024.

\bibitem[Hao et~al.(2023)Hao, Gu, Ma, Hong, Wang, Wang, and
  Hu]{hao2023reasoning}
Hao, S., Gu, Y., Ma, H., Hong, J., Wang, Z., Wang, D., and Hu, Z.
\newblock Reasoning with language model is planning with world model.
\newblock In \emph{Proceedings of the 2023 Conference on Empirical Methods in
  Natural Language Processing}, pp.\  8154--8173, 2023.

\bibitem[Hendrycks et~al.(2021)Hendrycks, Basart, Kadavath, Mazeika, Arora,
  Guo, Burns, Puranik, He, Song, et~al.]{hendrycks2021measuring}
Hendrycks, D., Basart, S., Kadavath, S., Mazeika, M., Arora, A., Guo, E.,
  Burns, C., Puranik, S., He, H., Song, D., et~al.
\newblock Measuring coding challenge competence with apps.
\newblock \emph{arXiv preprint arXiv:2105.09938}, 2021.

\bibitem[Hui et~al.(2024)Hui, Yang, Cui, Yang, Liu, Zhang, Liu, Zhang, Yu, Lu,
  et~al.]{hui2024qwen2}
Hui, B., Yang, J., Cui, Z., Yang, J., Liu, D., Zhang, L., Liu, T., Zhang, J.,
  Yu, B., Lu, K., et~al.
\newblock Qwen2. 5-coder technical report.
\newblock \emph{arXiv preprint arXiv:2409.12186}, 2024.

\bibitem[Jiang et~al.(2024{\natexlab{a}})Jiang, Wang, Shen, Kim, and
  Kim]{jiang2024survey}
Jiang, J., Wang, F., Shen, J., Kim, S., and Kim, S.
\newblock A survey on large language models for code generation.
\newblock \emph{arXiv preprint arXiv:2406.00515}, 2024{\natexlab{a}}.

\bibitem[Jiang et~al.(2024{\natexlab{b}})Jiang, Dong, Wang, Fang, Shang, Li,
  Jin, and Jiao]{jiang2024self}
Jiang, X., Dong, Y., Wang, L., Fang, Z., Shang, Q., Li, G., Jin, Z., and Jiao,
  W.
\newblock Self-planning code generation with large language models.
\newblock \emph{ACM Transactions on Software Engineering and Methodology},
  33\penalty0 (7):\penalty0 1--30, 2024{\natexlab{b}}.

\bibitem[Kurant et~al.(2010)Kurant, Markopoulou, and Thiran]{kurant2010bias}
Kurant, M., Markopoulou, A., and Thiran, P.
\newblock On the bias of {BFS} (breadth first search).
\newblock In \emph{2010 22nd International Teletraffic Congress (LTC 22)}, pp.\
   1--8. IEEE, 2010.

\bibitem[Kwon(2025)]{kwon2025vllm}
Kwon, W.
\newblock \emph{vLLM: An Efficient Inference Engine for Large Language Models}.
\newblock PhD thesis, University of California, Berkeley, 2025.

\bibitem[Li \& Murr(2024)Li and Murr]{li2024humaneval}
Li, D. and Murr, L.
\newblock Humaneval on latest {GPT} models--2024.
\newblock \emph{arXiv preprint arXiv:2402.14852}, 2024.

\bibitem[Li et~al.(2025{\natexlab{a}})Li, Cao, Cao, Li, Tan, Keutzer, Xing,
  Gonzalez, and Stoica]{li2025s}
Li, D., Cao, S., Cao, C., Li, X., Tan, S., Keutzer, K., Xing, J., Gonzalez,
  J.~E., and Stoica, I.
\newblock S*: Test time scaling for code generation.
\newblock \emph{arXiv preprint arXiv:2502.14382}, 2025{\natexlab{a}}.

\bibitem[Li et~al.(2025{\natexlab{b}})Li, Le, Zhou, Xiong, Savarese, and
  Sahoo]{li2025codetree}
Li, J., Le, H., Zhou, Y., Xiong, C., Savarese, S., and Sahoo, D.
\newblock Codetree: Agent-guided tree search for code generation with large
  language models.
\newblock In \emph{Proceedings of the 2025 Conference of the Nations of the
  Americas Chapter of the Association for Computational Linguistics: Human
  Language Technologies (Volume 1: Long Papers)}, pp.\  3711--3726,
  2025{\natexlab{b}}.

\bibitem[Li et~al.(2025{\natexlab{c}})Li, Dai, Li, Zhang, Wang, Tang, and
  Yu]{li2025codeprm}
Li, Q., Dai, X., Li, X., Zhang, W., Wang, Y., Tang, R., and Yu, Y.
\newblock Codeprm: Execution feedback-enhanced process reward model for code
  generation.
\newblock In \emph{Findings of the Association for Computational Linguistics:
  ACL 2025}, pp.\  8169--8182, 2025{\natexlab{c}}.

\bibitem[Li et~al.(2025{\natexlab{d}})Li, Dai, Liu, Li, Wang, Tang, Yu, and
  Zhang]{li2025atgen}
Li, Q., Dai, X., Liu, W., Li, X., Wang, Y., Tang, R., Yu, Y., and Zhang, W.
\newblock Atgen: Adversarial reinforcement learning for test case generation.
\newblock \emph{arXiv preprint arXiv:2510.14635}, 2025{\natexlab{d}}.

\bibitem[Li et~al.(2025{\natexlab{e}})Li, Xia, Dai, Du, Liu, Wang, Tang, Yu,
  and Zhang]{li2025rethinkmcts}
Li, Q., Xia, W., Dai, X., Du, K., Liu, W., Wang, Y., Tang, R., Yu, Y., and
  Zhang, W.
\newblock Rethinkmcts: Refining erroneous thoughts in monte carlo tree search
  for code generation.
\newblock In \emph{Proceedings of the 2025 Conference on Empirical Methods in
  Natural Language Processing}, pp.\  8103--8121, 2025{\natexlab{e}}.

\bibitem[Li et~al.(2023{\natexlab{a}})Li, Allal, Zi, Muennighoff, Kocetkov,
  Mou, Marone, Akiki, Li, Chim, et~al.]{li2023starcoder}
Li, R., Allal, L.~B., Zi, Y., Muennighoff, N., Kocetkov, D., Mou, C., Marone,
  M., Akiki, C., Li, J., Chim, J., et~al.
\newblock Starcoder: may the source be with you!
\newblock \emph{arXiv preprint arXiv:2305.06161}, 2023{\natexlab{a}}.

\bibitem[Li et~al.(2023{\natexlab{b}})Li, Fu, Zhang, Huang, Sun, Lyu, Liu, Jin,
  and Li]{li2023taco}
Li, R., Fu, J., Zhang, B.-W., Huang, T., Sun, Z., Lyu, C., Liu, G., Jin, Z.,
  and Li, G.
\newblock Taco: Topics in algorithmic code generation dataset.
\newblock \emph{arXiv preprint arXiv:2312.14852}, 2023{\natexlab{b}}.

\bibitem[Li et~al.(2022)Li, Choi, Chung, Kushman, Schrittwieser, Leblond,
  Eccles, Keeling, Gimeno, Dal~Lago, et~al.]{li2022competition}
Li, Y., Choi, D., Chung, J., Kushman, N., Schrittwieser, J., Leblond, R.,
  Eccles, T., Keeling, J., Gimeno, F., Dal~Lago, A., et~al.
\newblock Competition-level code generation with alphacode.
\newblock \emph{Science}, 378\penalty0 (6624):\penalty0 1092--1097, 2022.

\bibitem[Li et~al.(2025{\natexlab{f}})Li, Zhang, Zhang, Zhang, Liu, Yao, Xu,
  Zheng, Wang, Chen, et~al.]{li2025system}
Li, Z.-Z., Zhang, D., Zhang, M.-L., Zhang, J., Liu, Z., Yao, Y., Xu, H., Zheng,
  J., Wang, P.-J., Chen, X., et~al.
\newblock From system 1 to system 2: A survey of reasoning large language
  models.
\newblock \emph{arXiv preprint arXiv:2502.17419}, 2025{\natexlab{f}}.

\bibitem[Liu et~al.(2025)Liu, Mei, Lin, Xue, Wang, Xu, Wu, Zhang, Lin, Dong,
  et~al.]{liu2025deepseek}
Liu, A., Mei, A., Lin, B., Xue, B., Wang, B., Xu, B., Wu, B., Zhang, B., Lin,
  C., Dong, C., et~al.
\newblock Deepseek-v3. 2: Pushing the frontier of open large language models.
\newblock \emph{arXiv preprint arXiv:2512.02556}, 2025.

\bibitem[Lotov \& Miettinen(2008)Lotov and Miettinen]{lotov2008visualizing}
Lotov, A.~V. and Miettinen, K.
\newblock Visualizing the pareto frontier.
\newblock In \emph{Multiobjective {O}ptimization: {I}nteractive and
  {E}volutionary {A}pproaches}, pp.\  213--243. Springer, 2008.

\bibitem[Lozhkov et~al.(2024)Lozhkov, Li, Allal, Cassano, Lamy-Poirier, Tazi,
  Tang, Pykhtar, Liu, Wei, et~al.]{lozhkov2024starcoder}
Lozhkov, A., Li, R., Allal, L.~B., Cassano, F., Lamy-Poirier, J., Tazi, N.,
  Tang, A., Pykhtar, D., Liu, J., Wei, Y., et~al.
\newblock Starcoder 2 and the stack v2: The next generation.
\newblock \emph{arXiv preprint arXiv:2402.19173}, 2024.

\bibitem[Lyu et~al.(2025)Lyu, Huang, Deng, Hoi, and An]{lyu2025let}
Lyu, Z., Huang, J., Deng, Y., Hoi, S., and An, B.
\newblock Let's revise step-by-step: A unified local search framework for code
  generation with {LLMs}.
\newblock \emph{arXiv preprint arXiv:2508.07434}, 2025.

\bibitem[Mirzayanov et~al.(2020)Mirzayanov, Pavlova, MAVRIN, Melnikov,
  Plotnikov, Parfenov, and Stankevich]{mirzayanov2020codeforces}
Mirzayanov, M., Pavlova, O., MAVRIN, P., Melnikov, R., Plotnikov, A., Parfenov,
  V., and Stankevich, A.
\newblock Codeforces as an educational platform for learning programming in
  digitalization.
\newblock \emph{Olympiads in Informatics}, 14\penalty0 (133-142):\penalty0 14,
  2020.

\bibitem[Muennighoff et~al.(2025)Muennighoff, Yang, Shi, Li, Fei-Fei,
  Hajishirzi, Zettlemoyer, Liang, Cand{\`e}s, and Hashimoto]{muennighoff2025s1}
Muennighoff, N., Yang, Z., Shi, W., Li, X.~L., Fei-Fei, L., Hajishirzi, H.,
  Zettlemoyer, L., Liang, P., Cand{\`e}s, E., and Hashimoto, T.~B.
\newblock s1: Simple test-time scaling.
\newblock In \emph{Proceedings of the 2025 Conference on Empirical Methods in
  Natural Language Processing}, pp.\  20286--20332, 2025.

\bibitem[Ni et~al.(2023)Ni, Iyer, Radev, Stoyanov, Yih, Wang, and
  Lin]{ni2023lever}
Ni, A., Iyer, S., Radev, D., Stoyanov, V., Yih, W.-t., Wang, S., and Lin, X.~V.
\newblock Lever: Learning to verify language-to-code generation with execution.
\newblock In \emph{International Conference on Machine Learning}, pp.\
  26106--26128. PMLR, 2023.

\bibitem[Odeh et~al.(2024)Odeh, Odeh, and Mohammed]{odeh2024comparative}
Odeh, A., Odeh, N., and Mohammed, A.~S.
\newblock A comparative review of ai techniques for automated code generation
  in software development: advancements, challenges, and future directions.
\newblock \emph{TEM Journal}, 13\penalty0 (1):\penalty0 726, 2024.

\bibitem[Paul et~al.(2024)Paul, Zhu, and Bayley]{paul2024benchmarks}
Paul, D.~G., Zhu, H., and Bayley, I.
\newblock Benchmarks and metrics for evaluations of code generation: A critical
  review.
\newblock In \emph{2024 IEEE International Conference on Artificial
  Intelligence Testing (AITest)}, pp.\  87--94. IEEE, 2024.

\bibitem[Princis et~al.(2025)Princis, Sharma, and David]{princis2025treecoder}
Princis, H., Sharma, A., and David, C.
\newblock Treecoder: Systematic exploration and optimisation of decoding and
  constraints for {LLM} code generation.
\newblock \emph{arXiv preprint arXiv:2511.22277}, 2025.

\bibitem[Shao et~al.(2025)Shao, Zhang, Cheng, and Zhang]{shao2025decision}
Shao, T., Zhang, K., Cheng, K., and Zhang, H.
\newblock A decision-making framework using mcts as a hierarchical task network
  and deep learning connector.
\newblock \emph{Science Progress}, 108\penalty0 (4):\penalty0
  00368504251386308, 2025.

\bibitem[Shinn et~al.(2023)Shinn, Cassano, Gopinath, Narasimhan, and
  Yao]{shinn2023reflexion}
Shinn, N., Cassano, F., Gopinath, A., Narasimhan, K., and Yao, S.
\newblock Reflexion: Language agents with verbal reinforcement learning.
\newblock \emph{Advances in {N}eural {I}nformation {P}rocessing {S}ystems},
  36:\penalty0 8634--8652, 2023.

\bibitem[Silver et~al.(2016)Silver, Huang, Maddison, Guez, Sifre, Van
  Den~Driessche, Schrittwieser, Antonoglou, Panneershelvam, Lanctot,
  et~al.]{silver2016mastering}
Silver, D., Huang, A., Maddison, C.~J., Guez, A., Sifre, L., Van Den~Driessche,
  G., Schrittwieser, J., Antonoglou, I., Panneershelvam, V., Lanctot, M.,
  et~al.
\newblock Mastering the game of go with deep neural networks and tree search.
\newblock \emph{{N}ature}, 529\penalty0 (7587):\penalty0 484--489, 2016.

\bibitem[Silver et~al.(2017)Silver, Hubert, Schrittwieser, Antonoglou, Lai,
  Guez, Lanctot, Sifre, Kumaran, Graepel, et~al.]{silver2017mastering}
Silver, D., Hubert, T., Schrittwieser, J., Antonoglou, I., Lai, M., Guez, A.,
  Lanctot, M., Sifre, L., Kumaran, D., Graepel, T., et~al.
\newblock Mastering chess and shogi by self-play with a general reinforcement
  learning algorithm.
\newblock \emph{arXiv preprint arXiv:1712.01815}, 2017.

\bibitem[Snell et~al.(2024)Snell, Lee, Xu, and Kumar]{snell2024scaling}
Snell, C., Lee, J., Xu, K., and Kumar, A.
\newblock Scaling {LLM} test-time compute optimally can be more effective than
  scaling model parameters.
\newblock \emph{arXiv preprint arXiv:2408.03314}, 2024.

\bibitem[Wang et~al.(2025)Wang, Zhang, Feng, Li, Sun, Liu, and
  Peng]{wang2025teaching}
Wang, C., Zhang, J., Feng, Y., Li, T., Sun, W., Liu, Y., and Peng, X.
\newblock Teaching code {LLMs} to use autocompletion tools in repository-level
  code generation.
\newblock \emph{ACM Transactions on Software Engineering and Methodology},
  34\penalty0 (7):\penalty0 1--27, 2025.

\bibitem[Wang et~al.(2024)Wang, Cassano, Wu, Bai, Song, Nath, Han, Hendryx,
  Yue, and Zhang]{wang2024planning}
Wang, E., Cassano, F., Wu, C., Bai, Y., Song, W., Nath, V., Han, Z., Hendryx,
  S., Yue, S., and Zhang, H.
\newblock Planning in natural language improves {LLM} search for code
  generation.
\newblock \emph{arXiv preprint arXiv:2409.03733}, 2024.

\bibitem[Wang(2025)]{wang2025reward}
Wang, X.
\newblock Reward-centered rest-mcts: A robust decision-making framework for
  robotic manipulation in high uncertainty environments.
\newblock \emph{arXiv preprint arXiv:2503.05226}, 2025.

\bibitem[Wei et~al.(2022)Wei, Wang, Schuurmans, Bosma, Xia, Chi, Le, Zhou,
  et~al.]{wei2022chain}
Wei, J., Wang, X., Schuurmans, D., Bosma, M., Xia, F., Chi, E., Le, Q.~V.,
  Zhou, D., et~al.
\newblock Chain-of-thought prompting elicits reasoning in large language
  models.
\newblock \emph{Advances in {N}eural {I}nformation {P}rocessing {S}ystems},
  35:\penalty0 24824--24837, 2022.

\bibitem[Wu et~al.(2025)Wu, Chen, Lin, Zhan, Li, Kuang, and
  Wu]{wu2025personalized}
Wu, T., Chen, J., Lin, W., Zhan, J., Li, M., Kuang, K., and Wu, F.
\newblock Personalized distractor generation via mcts-guided reasoning
  reconstruction.
\newblock \emph{arXiv preprint arXiv:2508.11184}, 2025.

\bibitem[Xiang et~al.(2025)Xiang, Snell, Gandhi, Albalak, Singh, Blagden,
  Phung, Rafailov, Lile, Mahan, et~al.]{xiang2025towards}
Xiang, V., Snell, C., Gandhi, K., Albalak, A., Singh, A., Blagden, C., Phung,
  D., Rafailov, R., Lile, N., Mahan, D., et~al.
\newblock Towards system 2 reasoning in {LLMs}: Learning how to think with meta
  chain-of-thought.
\newblock \emph{arXiv preprint arXiv:2501.04682}, 2025.

\bibitem[Yang et~al.(2025{\natexlab{a}})Yang, Li, Yang, Zhang, Hui, Zheng, Yu,
  Gao, Huang, Lv, et~al.]{yang2025qwen3}
Yang, A., Li, A., Yang, B., Zhang, B., Hui, B., Zheng, B., Yu, B., Gao, C.,
  Huang, C., Lv, C., et~al.
\newblock Qwen3 technical report.
\newblock \emph{arXiv preprint arXiv:2505.09388}, 2025{\natexlab{a}}.

\bibitem[Yang et~al.(2025{\natexlab{b}})Yang, Chen, Gao, Li, Hu, Liu, and
  Xia]{yang2025empirical}
Yang, Z., Chen, S., Gao, C., Li, Z., Hu, X., Liu, K., and Xia, X.
\newblock An empirical study of retrieval-augmented code generation: Challenges
  and opportunities.
\newblock \emph{ACM Transactions on Software Engineering and Methodology},
  2025{\natexlab{b}}.

\bibitem[Yao et~al.(2023)Yao, Yu, Zhao, Shafran, Griffiths, Cao, and
  Narasimhan]{yao2023tree}
Yao, S., Yu, D., Zhao, J., Shafran, I., Griffiths, T., Cao, Y., and Narasimhan,
  K.
\newblock Tree of thoughts: Deliberate problem solving with large language
  models.
\newblock \emph{Advances in {N}eural {I}nformation {P}rocessing {S}ystems},
  36:\penalty0 11809--11822, 2023.

\bibitem[Zeng et~al.(2025)Zeng, Cheng, Yin, Zhou, and Qiu]{zeng2025revisiting}
Zeng, Z., Cheng, Q., Yin, Z., Zhou, Y., and Qiu, X.
\newblock Revisiting the test-time scaling of o1-like models: Do they truly
  possess test-time scaling capabilities?
\newblock \emph{arXiv preprint arXiv:2502.12215}, 2025.

\bibitem[Zhang et~al.(2025)Zhang, Lyu, Sun, Wang, Zhang, Hua, Wu, Guo, Wang,
  Muennighoff, et~al.]{zhang2025survey}
Zhang, Q., Lyu, F., Sun, Z., Wang, L., Zhang, W., Hua, W., Wu, H., Guo, Z.,
  Wang, Y., Muennighoff, N., et~al.
\newblock A survey on test-time scaling in large language models: What, how,
  where, and how well?
\newblock \emph{arXiv preprint arXiv:2503.24235}, 2025.

\bibitem[Zhang et~al.(2023)Zhang, Chen, Shen, Ding, Tenenbaum, and
  Gan]{zhang2023planning}
Zhang, S., Chen, Z., Shen, Y., Ding, M., Tenenbaum, J.~B., and Gan, C.
\newblock Planning with large language models for code generation.
\newblock \emph{arXiv preprint arXiv:2303.05510}, 2023.

\bibitem[Zhong et~al.(2024)Zhong, Wang, and Shang]{zhong2024ldb}
Zhong, L., Wang, Z., and Shang, J.
\newblock Ldb: A large language model debugger via verifying runtime execution
  step-by-step.
\newblock \emph{arXiv preprint arXiv:2402.16906}, 2024.

\bibitem[Zhou et~al.(2023)Zhou, Yan, Shlapentokh-Rothman, Wang, and
  Wang]{zhou2023language}
Zhou, A., Yan, K., Shlapentokh-Rothman, M., Wang, H., and Wang, Y.-X.
\newblock Language agent tree search unifies reasoning acting and planning in
  language models.
\newblock \emph{arXiv preprint arXiv:2310.04406}, 2023.

\end{thebibliography}
\bibliographystyle{icml2026}

\newpage
\appendix
\onecolumn
\section{Introduction of Monte Carlo Tree Search}
Monte Carlo Tree Search (MCTS) is a heuristic search algorithm that balances exploration and exploitation to navigate complex decision spaces, achieving significant success in sequential decision-making tasks \citep{silver2016mastering, shao2025decision, han2024value, wang2025reward}. MCTS iteratively builds a search tree where nodes represent states and edges represent actions. The process generally consists of four phases: \textit{Selection} (select a child node that maximize a tree policy), \textit{Expansion} (one or more child nodes are added to represent potential future states), \textit{Simulation} (simulate until a terminal state and get reward), and \textit{Backpropagation} (the reward is propagated back up the tree to the root and update the values of all traversed nodes).
In the context of reasoning or generation tasks, MCTS allows the model to look ahead and evaluate partial solutions~\citep{chaffin2022ppl, wu2025personalized}, guiding the generation process toward higher-quality outcomes compared to greedy decoding methods.

\section{Details of Baselines}
\label{app:baselines}

To evaluate the effectiveness of \our{}, we compare it against a comprehensive set of baselines ranging from direct prompting to advanced tree-search methods. The specific implementation details for each baseline are as follows:

\begin{itemize}
    \item \textbf{Direct Synthesis (Base \& Best-of-N):} The \textit{Base} method employs standard zero-shot prompting to generate a single solution. \textit{Best-of-N} scales this by sampling $N=16$ independent candidates ($T=0.7$) and selecting the best one based on public test cases, serving as a strong sampling-based baseline without lookahead search.

    \item \textbf{PG-TD \citep{zhang2023planning}:} This method effectively utilizes MCTS to perform lookahead search at the token level. It explores the probability space of initial tokens to identify promising prefixes, which are then deterministically completed into full code solutions, aiming to guide generation through high-likelihood token trajectories.

    \item \textbf{Tree of Thoughts (ToT) \citep{yao2023tree}:} We implement a structured, two-layer search tree adapted for code generation. The first layer expands natural language \textit{plans} or algorithmic sketches, while the second layer generates multiple concrete \textit{code implementations} for each plan. This hierarchical approach decouples high-level reasoning from low-level implementation details.

    \item \textbf{Language Agent Tree Search (LATS) \citep{zhou2023language}:} LATS unifies reasoning and planning by searching directly in the \textit{code space} with an integrated reflection mechanism. Upon encountering a failure during simulation, the agent generates verbal self-reflection based on the error feedback; this reflection is incorporated into the context to guide the value estimation and selection of subsequent nodes.

    \item \textbf{MCTS-Thought:} This is a search baseline that performs MCTS over Chain-of-Thought (CoT) reasoning steps rather than raw code. In the simulation phase, it synthesizes full code based on the current thought trajectory; crucially, execution feedback from these generated codes (e.g., test results) is added to the context to inform the expansion of the next reasoning step in a multi-turn manner.

    \item \textbf{RethinkMCTS \citep{li2025rethinkmcts}:} Similar to MCTS-Thought, this method searches in the thought space but introduces a specific \textit{repair mechanism}. Instead of simply discarding nodes that lead to incorrect code, it utilizes fine-grained execution feedback to explicitly "rethink" and refine erroneous thought nodes, allowing the search to recover from early reasoning mistakes without restarting.
\end{itemize}

\section{Additional Experiments}

\subsection{Empirical Validation of Pseudo-Correctness: Is the Solution Generated but Overlooked?}
\label{app:emp_pesudo_correct}
\paragraph{Motivation.} A core premise of \our{} is that existing search methods (e.g., MCTS-Thought) suffer from pseudo-correctness: they successfully generate correct solutions during the search process, but fail to identify them due to the sparsity of public test cases (typically $|\mathcal{T}_{\text{pub}}|=5$.To validate this hypothesis, we conducted a controlled experiment to decouple ``generation capability'' from ``verification capability''.

\paragraph{Experimental Setup.} We employed the MCTS-Thought baseline with the Qwen3-4B-Instruct model. Instead of changing the search algorithm, we only varied the verification environment used during the selection and backpropagation phases. We compared three settings:
\begin{itemize}
    \item \textbf{Original (5 Tests):} Standard setting with 5 public tests.
    \item \textbf{Half Tests (50\%):} Using 50\% of the hidden test suite as visible constraints.
    \item \textbf{Oracle (All Tests):} Using the full hidden test suite (simulate perfect verification).
\end{itemize}

\begin{figure}[h]
    \centering
    \includegraphics[width=0.42\textwidth]{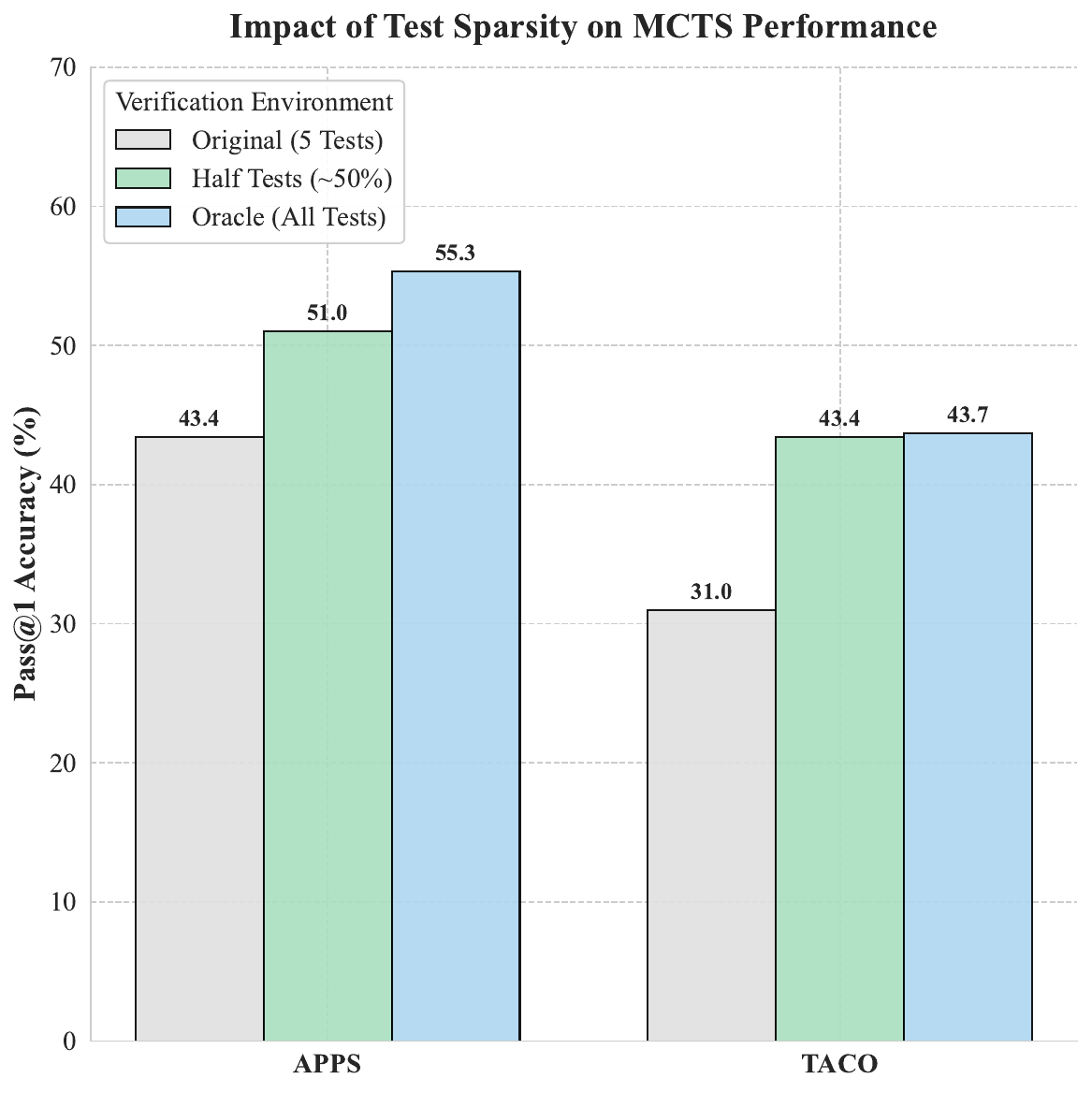}
    \caption{\textbf{Empirical Validation of Pseudo-Correctness.} Comparison of MCTS-Thought performance under varying verification environments (Original, Half-Hidden, Oracle). The significant performance gap between the standard setting (5 Tests) and the Oracle setting confirms the prevalence of pseudo-correctness: robust solutions are successfully generated but are filtered out due to the sparsity of public tests.}
    \label{fig:bar}
    \vspace{-10pt}
\end{figure}

\paragraph{Results and Analysis.} As shown in Figure ~\ref{fig:bar}, the performance of the exact same search algorithm improves dramatically as the verification environment becomes stricter.
\begin{itemize}
    \item On APPS, increasing test coverage boosts Pass@1 from 43.44\% to 55.33\% (+11.89\%).
    \item On TACO, the gain is even more pronounced, jumping from 31.00\% to 43.67\% (+12.67\%).
\end{itemize}

These results provide compelling evidence that \textbf{pseudo-correctness is the primary bottleneck}. The search space already contains robust solutions (as evidenced by the high performance in the Oracle setting), but the standard sparse filter allows fragile solutions to ``survive'' and crowd them out. This justifies the design of \our{}: since we cannot access the Oracle tests in practice, we must actively generate adversarial tests to approximate this strict verification environment.

\subsection{Hard Re-ranking Outperforms Soft Penalties.}

We investigate the optimal integration of adversarial feedback by comparing \textit{Direct Penalization} (continuous value subtraction) with our \textit{Test-based Re-ranking} (discrete filtering). Table~\ref{tab:penalty_vs_rerank} demonstrates that re-ranking consistently yields superior performance. This result indicates that adversarial tests are most effective when utilized as hard constraints rather than soft regularizers. While direct penalization introduces noise due to arbitrary scaling (e.g., equating core logic failures with minor edge cases), re-ranking aligns with the binary nature of unit testing, strictly filtering brittle solutions that fail to survive the hostile environment.

\begin{table}[t]
    \centering
    \caption{Signal Utilization Strategy. Comparison of using adversarial signals as a continuous penalty (\textit{Direct Penalty}) versus a discrete filter (\textit{Re-ranking}). The \textit{Re-ranking} strategy consistently outperforms direct penalization, suggesting that adversarial tests are most effective when treated as hard constraints.}
    \label{tab:penalty_vs_rerank}
    \resizebox{0.48\textwidth}{!}{
    \begin{tabular}{l|ccc|ccc}
        \toprule
        \multirow{2}{*}{\textbf{Method}} & \multicolumn{3}{c|}{\textbf{APPS (Pass@1)}} & \multicolumn{3}{c}{\textbf{TACO (Pass@1)}} \\
        \cmidrule(lr){2-4} \cmidrule(lr){5-7}
         & \textbf{Intro.} & \textbf{Inter.} & \textbf{Comp.} & \textbf{Easy} & \textbf{Medium} & \textbf{Hard} \\
        \midrule
        Direct Penalty & 57.0 & 45.0 & \textbf{36.0} & 49.0 & 32.0 & 17.0 \\
        Re-ranking & \textbf{61.0} & \textbf{52.0} & \textbf{36.0} & \textbf{55.0} & \textbf{37.0} & \textbf{20.0} \\
        \midrule
        \textit{Improvement} & \textit{+4.0} & \textit{+7.0} & \textit{+0.0} & \textit{+6.0} & \textit{+5.0} & \textit{+3.0} \\
        \bottomrule
    \end{tabular}
    }
    \vspace{-10pt}
\end{table}

\begin{figure}[htbp] 
    \centering
    \begin{subfigure}[t]{0.48\columnwidth}
           \centering
           \includegraphics[width=\columnwidth]{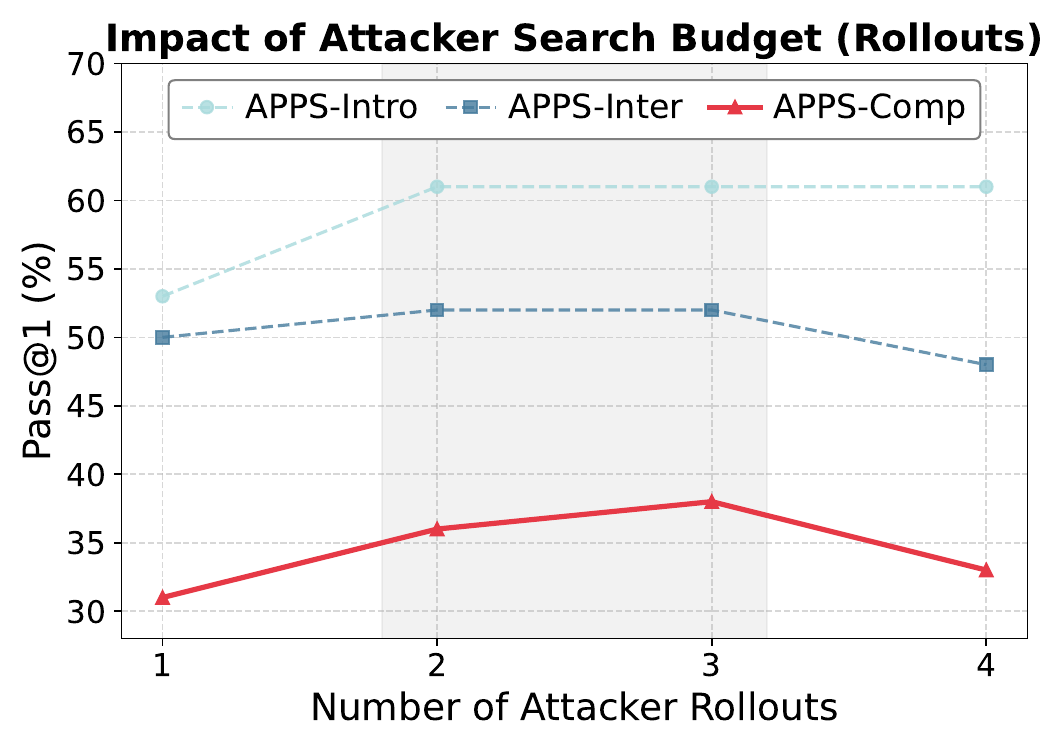}
            \caption{APPS}
    \end{subfigure}
    \begin{subfigure}[t]{0.48\columnwidth}
           \centering
           \includegraphics[width=\columnwidth]{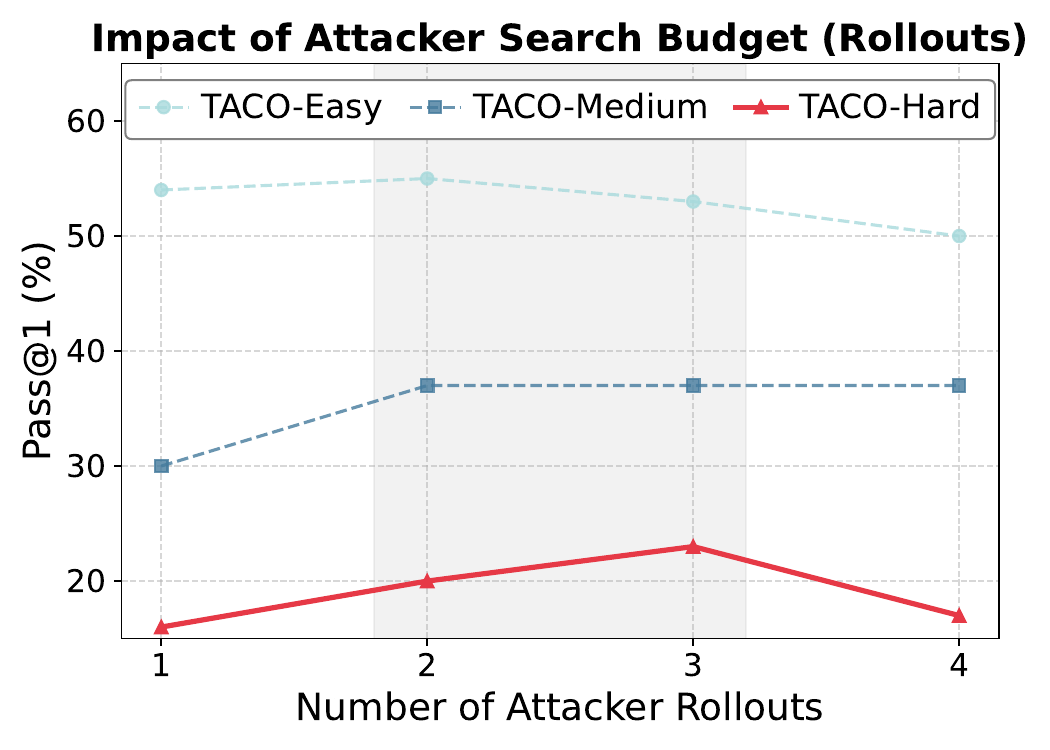}
            \caption{TACO}
    \end{subfigure}
    \centering
    \caption{Attacker rollout scaling. Varying the Attacker’s number of rollouts shows an inverted-U trend on both APPS and TACO: moderate budgets improve Pass@1 across difficulty splits, while larger budgets can saturate or degrade performance.}
    \label{fig:rollout_scaling}
\end{figure}

\subsection{Impact of Attacker Search Budget: More is Not Always Better.}
A natural question is whether allocating more test-time compute to the Attacker (i.e., more rollouts) monotonically improves the Solver by producing increasingly challenging adversarial tests. To study this, we vary the Attacker rollout limit from 1 to 4 while keeping other components fixed, and evaluate downstream Solver performance on APPS.

Figure~\ref{fig:rollout_scaling} reveals a clear inverted-U behavior: increasing rollouts from 1 to 2 yields a substantial gain (e.g., Pass@1 on APPS-Competition improves from 31\% to 36\%), indicating that a minimal search budget is necessary for MCTS to move beyond trivial cases and discover discriminative corner tests. However, further increasing the budget brings diminishing returns and can even hurt performance. We attribute this to a practical trade-off: a more aggressive search may over-optimize the divergence objective and produce tests that are less reliable as supervision signals (e.g., brittle constraints or borderline-valid corner cases), which can distort screening and re-ranking. Based on this observation, we use a moderate rollout budget ($N=2$) in \our{} as it provides a favorable balance between effectiveness and efficiency.

\section{Prompts}
\label{app:prompts}

In this section, we present the prompts used when an LLM acts as an agent to perform various operations.
\subsection{Solver Expansion Prompt}
We present the prompt we use to instruct the LLM to sample code generation thoughts based on previous thoughts and possible previous code execution feedback.
\begin{promptbox}{Solver Expansion Prompt}
\small\ttfamily

\role{System}

You are a code reasoning agent.

Your task is to generate one plan step of reasoning for solving the programming problem.

Strictly output ONLY the thought content without any prefix or formatting.

Be concise but comprehensive (2-3 sentences maximum).

\vspace{0.5em} \hrule \vspace{0.5em} 

\role{User\ Prompt}

\placeholder{Problem}

\placeholder{Previous Thoughts}

\placeholder{Previous Code Implementation}

\placeholder{Previous Execution Results}

Provide the next logical thought to solve the problem. If previous thoughts exist, build upon them to offer deeper insight.

Ensure the reasoning leads to code that handles edge cases and prevents the current error from recurring.
\end{promptbox}

\subsection{Solver Code Generation Prompt}
We present the prompt used to instruct the LLM to generate code based on previous thoughts.

\begin{promptbox}{Solver Code Generation Prompt}
\small\ttfamily

\role{System}

You are a code generator.

\vspace{0.5em} \hrule \vspace{0.5em} 

\role{User\ Prompt}

\placeholder{Problem Description}

\placeholder{Previous Thoughts}

Based on the above problem description and the reasoning process (thoughts), please generate the complete Python code to solve the problem.

The solution should contain the complete program including all the imports.

Generate the code ONLY. No other explanation or words attached!

Please wrap the solution into \texttt{```python ... ```} format.

\end{promptbox}


\subsection{Attacker Expansion Prompt}
We present the prompt we use to instruct the LLM to sample test generation thoughts based on previous thoughts and possible previous test execution feedback on the code pool.

\begin{promptbox}{Attacker Expansion Prompt}
\small\ttfamily

\role{System}

You are a test thought designer specializing in finding bugs through systematic testing approaches.

Your task is to propose a distinct testing thought that could reveal bugs in code implementations.

Output your strategy directly.

Keep the thought concise (1-2 sentences) and actionable.

\vspace{0.5em} \hrule \vspace{0.5em} 

\role{User\ Prompt}

\placeholder{Problem Description}

\placeholder{Previous Thoughts}

You should propose a MORE SPECIFIC sub-thought that refines the current approach.
Previous Execution Feedback:

\placeholder{feedback context}

Generate a distinct testing thought that could help find bugs or differences between code implementations.

Focus on:

1. Edge cases and boundary conditions

2. Special input patterns

3. Common algorithmic pitfalls for this type of problem

4. Input formats that often cause subtle bugs

DO NOT generate specific test inputs - only describe the STRATEGY/APPROACH.

Your thought:

\end{promptbox}

\subsection{Attacker Test Input Generation Prompt}
We present the prompt we use to instruct the LLM to generate test input following previous thoughts and based on current codes in the code pool.

\begin{promptbox}{Attacker Test Input Generation Prompt}
\small\ttfamily

\role{System}

You are a helpful AI Assistant that provides well-reasoned and detailed responses. You first think about the reasoning process as an internal monologue and then provide the user with the answer. Respond in the following format: <think>...</think><answer>...</answer>. Think concisely in 2-3 sentences.

\vspace{0.5em} \hrule \vspace{0.5em} 

\role{User\ Prompt}

\placeholder{Problem Description}

\placeholder{Previous Thoughts}

Current Code Implementations to Test:

\placeholder{codes}

Based on the testing thought above, generate ONE specific test input that:

1. Follows the thought approach

2. Is valid according to the problem description

3. Is likely to reveal differences between the code implementations OR expose bugs

The codes have already passed the following public tests (use this context to find DIFFERENT corner cases):

\placeholder{public tests}

Format your answer as a JSON object containing only the "input" key, enclosed in triple backticks ```json ```. For example:

<answer>

```json

\{"input": "[your generated test case input]"\}

```

</answer>

Important: Only provide the INPUT, not the expected output. We will execute the codes to compare their outputs.
\end{promptbox}

\subsection{Attacker Test Output Arbiter Prompt}
We present the prompt we use to instruct the LLM to determine the correct output from the diverse outputs from the code pool.

\begin{promptbox}{Attacker Test Output Arbiter Prompt}
\small\ttfamily
\role{User\ Prompt}

You are an expert programmer. Given a programming problem and test input, multiple code implementations produced different outputs. Determine which output is CORRECT.

Problem Description:

\placeholder{Problem Description}

Public Tests (Use as reference for correct behavior):

\placeholder{public tests}

Test Input:

\placeholder{Test Input}

Code Outputs:

\placeholder{Outputs}

Instructions:

1. Understand what the problem asks

2. Trace through the logic with the given test input 

3. Determine the CORRECT output and which code(s) produced it
    
Respond in the following format:

<reasoning>

Brief explanation (2-3 sentences max) of why this is the correct output.

</reasoning>

<correct\_output>

The correct output value

</correct\_output>

<correct\_codes\_id>

List of correct code indices, e.g., [1, 3] or [2]

</correct\_codes\_id>

\end{promptbox}

\section{Algorithm}
\label{ag:algs} %
We present the detailed procedure of \our{} in pseudocode in Algorithm~\ref{alg:advermcts}.

\begin{algorithm}[H]
\caption{The \textbf{\our{}} Inference-Time Search Procedure.}
\label{alg:advermcts}
\small{
\begin{algorithmic}[1]
\REQUIRE $P$: Problem description; $\mathcal{T}_{pub}$: Public tests; $N_{s\_iter}$: Iterations of Solver; $N_{a\_iter}$: Iterations of Attacker.
\STATE \textbf{Output:} Robust solution $C^*$

\STATE {\color{ForestGreen} \# Initialization}
\STATE Initialize Solver Tree $\mathcal{S}$ with root $s_0 = \{P\}$
\STATE Initialize Attacker Tree $\mathcal{A}$ with root $\sigma_0 = \{P, \emptyset\}$
\STATE Code Pool $\mathcal{C}_{pool} \leftarrow \emptyset$, Global Test Filter $\mathcal{T}_{global} \leftarrow \emptyset$

\FOR{$i \gets 1$ \textbf{to} $N_{s\_iter}$}
    \STATE {\color{ForestGreen} \# Phase 1: Solver Tree (Code Generation)}
    \STATE $s_{leaf} \leftarrow \texttt{\MakeUppercase{select}}(\mathcal{S})$
    \STATE $s_{new} \leftarrow \texttt{\MakeUppercase{expand}}(s_{leaf})$
    \STATE $C \leftarrow \texttt{\MakeUppercase{generate\_code}}(s_{new})$
    \STATE $r_{pub} \leftarrow \texttt{\MakeUppercase{evaluate}}(C, \mathcal{T}_{pub})$
    
    \STATE {\color{ForestGreen} \# Strict Admission}
    \IF{$r_{pub} = 1.0$ \textbf{and} $\texttt{\MakeUppercase{pass\_all}}(C, \mathcal{T}_{global})$}
        \STATE $\mathcal{C}_{pool} \leftarrow \mathcal{C}_{pool} \cup \{C\}$
        \IF{$|\mathcal{C}_{pool}| > K_{max}$}
            \STATE Remove oldest/lowest-score code from $\mathcal{C}_{pool}$
        \ENDIF
    \ENDIF

    \STATE {\color{ForestGreen} \# Phase 2: Attacker Tree (Vulnerability Discovery)}
    \IF{$|\mathcal{C}_{pool}| \ge 2$}
        \FOR{$i \gets 1$ \textbf{to} $N_{a\_iter}$}
            \STATE $\sigma_{leaf} \leftarrow \texttt{\MakeUppercase{select}}(\mathcal{A})$
            \STATE {\color{ForestGreen} \# Late-Binding: Thought + Code Pool $\to$ Concrete Test}
            \STATE $T_{candidates} \leftarrow \texttt{\MakeUppercase{gen\_test}}(\sigma_{leaf}, \mathcal{C}_{pool})$
            \STATE $t^* \leftarrow \arg\max_{t \in T_{candidates}} \texttt{\MakeUppercase{div\_score}}(t, \mathcal{C}_{pool})$
            \STATE $d^* \leftarrow \texttt{\MakeUppercase{div\_score}}(t^*, \mathcal{C}_{pool})$
    
            \IF{$d^* > \theta_{thresh}$}
                \STATE {\color{ForestGreen} \# Arbiter determines Ground Truth}
                \STATE $o^*, \mathcal{C}_{fail} \leftarrow \texttt{\MakeUppercase{arbiter}}(P, t^*, \texttt{\MakeUppercase{outputs}}(\mathcal{C}_{pool}, t^*))$
                
                \IF{$o^*$ is valid}
                    \STATE $\mathcal{T}_{global} \leftarrow \mathcal{T}_{global} \cup \{(t^*, o^*)\}$
                    \FORALL{$C \in \mathcal{C}_{fail}$}
                        \STATE Apply Penalty $-V$ to Solver nodes of $C$
                        \STATE Update Code Value $Q(s) \leftarrow Q(s) - V$
                    \ENDFOR
                \ENDIF
            \ENDIF
    
            \STATE $\texttt{\MakeUppercase{expand}}(\sigma_{leaf})$
            \STATE $\texttt{\MakeUppercase{backprop}}(\mathcal{A}, d^*)$
        \ENDFOR
    \ENDIF

    \STATE {\color{ForestGreen} \# Phase 3: Solver Update}
    \STATE $\texttt{\MakeUppercase{backprop}}(\mathcal{S}, r_{pub} - \text{AccumulatedPenalties})$
\ENDFOR

\STATE {\color{ForestGreen} \# Inference: Hierarchical Test-based Reranking}
\STATE Final Candidates $\mathcal{C}_{final} \leftarrow \texttt{\MakeUppercase{get\_all\_codes}}(\mathcal{S})$
\STATE \textbf{return} $\arg\max_{C \in \mathcal{C}_{final}} \langle \texttt{\MakeUppercase{pass}}(C, \mathcal{T}_{pub}), \texttt{\MakeUppercase{pass}}(C, \mathcal{T}_{global}) \rangle$
\end{algorithmic}
}
\end{algorithm}


\end{document}